\documentclass[a4paper,fleqn,usenatbib]{mnras}

\usepackage{amsmath}
\usepackage{amsfonts}
\usepackage{amsbsy}
\usepackage{amssymb}
\usepackage{amsbsy}
\usepackage{makecell}

\usepackage[T1]{fontenc}
\usepackage{ae,aecompl}

\usepackage{graphicx}
\usepackage{gensymb}
\usepackage{float}
\usepackage[utf8]{inputenc}
\usepackage{tensor}
\usepackage{bm}
\usepackage{enumerate}
\usepackage{comment}

\newcommand{\rmd}{{\rm d}}

\newcommand{\ey}{\hat{\bb{e}}_y}
\newcommand{\ez}{\hat{\bb{e}}_z}
\newcommand{\eb}{\hat{\bb{e}}_b}
\newcommand{\const}{{\rm const}}
\newcommand\mc[1]{\mathcal{#1}}

\newcommand{\pD}[2]{\frac{\partial #2}{\partial #1}}

\newcommand{\D}[2]{\frac{\rmd{#2}}{\rmd{#1}}}
\newcommand{\DD}[2]{\frac{\rmd^2 #2}{\rmd {#1}^2}}
\newcommand\bb[1]{\mbox{\boldmath{$#1$}}}
\newcommand\grad{\bb{\nabla}}
\newcommand\bcdot{\,\bb{\cdot}\,}
\newcommand\btimes{\,\bb{\times}\,}

\usepackage{xcolor}

\DeclareMathAlphabet\mathbfcal{OMS}{cmsy}{b}{n}

\defcitealias{riols17c}{RL18a}
\defcitealias{riols18}{RL18b}
\defcitealias{riols19b}{RL19}

\title[]{Gravito-turbulence and  dynamo in poorly ionised protostellar discs.~I.~Zero-net-flux case}
\author[]{
A.~Riols$^{1}$, W.~Xu$^{2}$, G.~Lesur$^{1}$, M.~W.~Kunz$^{2,3}$, H.~Latter$^{4}$
\\
$^{1}$Univ. Grenoble Alpes, CNRS, Institut de Planétologie et d’Astrophysique de Grenoble (IPAG), F-38000, Grenoble, France\\
$^{2}$Department of Astrophysical Sciences, 4 Ivy Lane, Peyton Hall, Princeton University, Princeton, NJ 08544, U.S.A.\\
$^{3}$Princeton Plasma Physics Laboratory, PO Box 451, Princeton, NJ 08543, U.S.A. \\
$^{4}$DAMTP, University of Cambridge, Centre for Mathematical Sciences,
Wilberforce Road, Cambridge CB3 0WA, U.K.}

\date{Accepted XXX. Received YYY; in original form ZZZ}

\pubyear{2020}

\begin{document}
\label{firstpage}
\pagerange{\pageref{firstpage}--\pageref{lastpage}}
\maketitle

\begin{abstract}
In their early stages, protoplanetary discs are sufficiently massive to undergo gravitational instability (GI). This instability is thought to be involved in mass accretion, planet formation via gas fragmentation, the generation of spiral density waves, and outbursts. A key and very recent area of research is the interaction between the GI and magnetic fields in young protoplanetary discs, in particular whether this instability is able to sustain a magnetic field via a dynamo. We conduct three-dimensional, stratified shearing-box simulations using two independent codes, {\tt PLUTO} and {\tt Athena++}, to characterise the GI dynamo in poorly ionised protostellar discs subject to ambipolar diffusion.  We find that the dynamo operates across a large range of ambipolar Elssaser number Am (which characterises the strength of ambipolar diffusion) and is particularly strong in the regime $\text{Am}=10$--$100$, with typical magnetic to thermal energy ratios of order unity. The dynamo is only weakly dependent on resolution (at least for $\text{Am} \lesssim 100$), box size, and cooling law. The magnetic field is produced by the combination of differential rotation and large-scale vertical roll motions associated with spiral density waves. Our results have direct implications for the dynamo process in young protoplanetary discs and possibly some regions of AGN discs.
\end{abstract}




\section{Introduction}
\label{sec_intro}

Magnetic fields should be ubiquitous in accretion discs and are likely to influence their dynamics and long-term evolution significantly. 
Disc magnetism is believed to facilitate the accretion process by enabling turbulent angular-momentum transport and by launching large-scale winds \citep{balbus03,wardle07,armitage15}. It is also possible that magnetic fields are responsible for the formation of large-scale structures, such as concentric rings \citep{kunz13,bai15,bethune17,riols19a,suriano19}, which are now widely observed in protoplanetary discs by ALMA and SPHERE \citep[e.g.,][]{alma15,sphere18,andrews20}. Despite their perceived importance, direct measurements of magnetic-field strengths in protoplanetary discs are rare. \citet{donati05} detected azimuthal fields at a strength of 1~kG in the inner part of the FU-Ori disc using the high-resolution spectropolarimeter ESPaDOnS. In the outer parts of discs, researchers have attempted to measure fields directly from the Zeeman splitting of CN lines \citep{vlemmings19}, and have set an upper limit of 20~mG on the toroidal field and 0.8~mG on the vertical field.  Indirect techniques using dust polarisation have detected magnetic fields with various intensity and morphologies  \citep{carrasco10,stephens14,goddi17}, but these observations remain disputed \citep{kataoka17,tazaki17}. Magnetic fields have been also recorded in meteorite samples, suggesting fields of a few tens of mG at $R > 1$ AU in the primitive solar system \citep{fu14,cournede15,fu20}. On the other hand, in active galactic nuclei (AGN) and binary discs there are some direct detections of a strong or moderate poloidal field \citep{martinvidal15}, but the existence of magnetic fields is mainly inferred from the jets associated with these objects \citep[see review by][]{spruit96}. 

A central but difficult question pertains to the origin of these magnetic fields and their sustenance against dissipative effects resulting from the microphysics of the plasma such as non-ideal magnetohydrodynamic (MHD) effects and/or turbulent diffusion. In discs around young stars, one possibility is that the residual field from the primordial nebulae is advected and amplified by the combined effects of gravitational contraction, rotation, and accretion.
{Indeed, simulations of magnetic star formation incorporating non-ideal MHD effects suggest a field strength ${\sim}0.1~{\rm G}$ in the central flux tubes of the `first protostellar core' \citep[e.g.,][]{DM01,KM10,masson16,XK21}, consistent with estimates for the protosolar magnetic field as deduced from meteoritic data \citep{LevySonett1978}. To what extent this magnetic flux is then retained by and transported through the protostellar accretion disc during the Class 0 and I phases depends upon the (unknown) level of turbulent transport and diffusivity in the disc \citep[e.g.,][]{guilet13,bai17,leung19}.}

Another possibility is that any weak seed magnetic field is amplified in the disc itself by a turbulent dynamo \citep[e.g.,][]{pudritz81,balbus98}. For nearly thirty years, the magneto-rotational instability \citep[MRI, ][]{balbus91,hawley95} has been considered as a primary source of turbulence and angular-momentum transport in astrophysical discs. Several studies have pointed out the existence of a sub-critical dynamo involving the MRI that exhibits regular reversals of the large-scale magnetic field, somewhat reminiscent of the `butterfly' diagram of the solar dynamo \citep{branden95,hawley96,lesur08}. However there is today a broad consensus that the MRI dynamo reliably survives only in the very inner parts of protoplanetary discs, within 0.1--1~AU where the temperature is high enough to thermally ionise the plasma \citep{gammie96,sano2000}.  Further away,  the ionisation fraction is too low and the dynamics is regulated by non-ideal MHD effects, such as Ohmic dissipation, the Hall effect, and ambipolar diffusion, which all tend to suppress any form of dynamo process \citep{fleming00, sano02,wardle12,bai13b,lesur14,bai15}. Even in the inner regions of protoplanetary discs and in more ionised objects like AGN discs, the viability of the MRI dynamo process is not guaranteed, because of the low magnetic Prandtl number that characterises the plasma in these environments \citep[see][for more discussions about this issue]{fromang07b,balbus08,kapyla11,meheut15,riols16c}. 

Another potential source of turbulence in accretion discs is the gravitational instability (GI). This instability is thought to manifest during the early phases of protoplanetary discs when their masses are large enough for self-gravity to be dynamically important. Depending on the efficiency of the cooling process, GI either triggers fragmentation of the disc or saturates in a `gravito-turbulent' state characterised by spiral density waves that are continuously generated and dissipated \citep{Gammie2001,rice03,rice06,durisen07}. Based on measurements of disc masses, recent surveys found that 50\% of Class 0 and 10-20\% of Class I sources could be unstable to GI \citep{tobin13, mann15}. Recent images of large-scale spirals arms in several protoplanetary discs (e.g.\ Elias 2-27, WaOph 6, MWC758) and more disordered `streamer' structures in FU Ori systems, have been attributed to GI activity \citep{liu16,dong16,perez16,huang18}. Gravitational instability is also believed to occur in AGN discs at distances of ${\sim}0.01$--$0.1~{\rm pc}$, though fragmentation (instead of gravito-turbulence) is expected because the cooling timescale is much shorter than the dynamical timescale \citep{lodato07}. In particular, this leads to the AGN disk truncation problem \citep{goodman03}.

Recently, the ability of GI motions to amplify and sustain a magnetic field has been investigated through  local \citep[][hereafter \citetalias{riols19b}]{riols19b} and global numerical simulations \citep{deng20}. These studies have shown that gravito-turbulence works as a dynamo itself, producing strong azimuthal fields that can reach nearly thermal strengths even in poorly ionised gas characterised by magnetic Reynolds number of order ${\sim}10$. The magnetic fields generated back-react on the flow significantly and tend to reduce the strength of the spiral density waves. The mechanism behind the amplification of the field, identified by \citetalias{riols19b}, involves the action of counter-rotating vertical rolls associated with the spiral density waves. These rolls stretch and fold azimuthal field lines ($\alpha$ effect), leading to the generation of a mean radial field, which in turn is sheared to produce toroidal fields ($\Omega$ effect). However, the GI-dynamo process has been examined in discs where magnetic dissipation results exclusively from Ohmic resistivity. In Class 0 discs, it is expected that other non-ideal effects, such as ambipolar diffusion, prevail in regions where GI may be active \citep[see][who have used a detailed chemical equilibrium model to compute diffusivities in nascent protostellar discs]{dapp12}. A natural question that arises is whether or not ambipolar diffusion changes the growth and saturation of the GI dynamo. In particular, is the amplification mechanism similar to that at work in discs regulated by Ohmic dissipation? The answer is non-trivial since there are major differences between the two diffusion processes: unlike Ohmic dissipation, ambipolar diffusion is non-linear, anisotropic, and does not allow reconnection of the field.

Another important question that has been only partially addressed \citep{riols17c} is the effect of a net vertical field threading the disc on the dynamo. Such a field, if sufficiently strong, could severely impede gravito-turbulence \citep{riols17c} and have direct consequences for the GI-dynamo process. In addition, such a vertical field is expected to facilitate the launching of a wind \citep{wk93,ferreira95,lesur2013,fromang13}.

In this paper, we focus on the zero-net-flux configuration (without a mean vertical field) and examine the effect of ambipolar diffusion on the gravito-turbulent dynamo. We performed 3D MHD stratified shearing-box simulations with self-gravity and vertical stratification using two different codes, {\tt PLUTO} and {\tt Athena++}. (The Hall effect and Ohmic dissipation are omitted.) Broad agreement between the code results is achieved, demonstrating their weak dependence on the implementation of self-gravity and on other numerical details (time-stepping methods, orbital advection algorithm, etc.). We explore different cooling prescriptions (namely, linear cooling and $T^4$ cooling) and consider a wide range of ambipolar Elsasser numbers (${\rm Am}$) from 1 to 200.  Given the size of the box necessary to capture GI and thus our limited resolution per $H$, it is not possible to probe the regime of yet larger ${\rm Am}$. In Paper II we examine the effect of a net vertical field on the dynamo behaviour by varying the midplane $\beta$ from $10^6$ to $10^3$.

We find that the gravito-turbulent dynamo in the presence of ambipolar diffusion is very similar to that described by \citetalias{riols19b}: there is no fundamental difference between the magnetic fields generated in the presence of ambipolar diffusion and those produced subject to Ohmic dissipation. Dynamo amplification is most efficient between $\text{Am}=30$ and $\text{Am}=100$, giving super-equipartition fields (with $\beta \sim 1$) that are predominantly toroidal and whose energy resides primarily at large scales. The dynamo weakens in the regime of large $\text{Am} \gtrsim 100$ and for sufficiently small Am. As in \citetalias{riols19b}, the field is amplified by poloidal velocity rolls associated with large-scale density waves. By using a new technique involving phase-folded simulation snapshots, we provide a clearer view of how gravitationally induced spiral arms amplify  magnetic fields.

The structure of the paper is as follows. In Section \ref{sec_model}, we describe the model used and the basic equations solved, as well as the numerical setup of our simulations. In Section \ref{sec_saturation}  we study the saturated state of the dynamo as a function of ${\rm Am}$ and compare the results amongst two different cooling laws. In Section \ref{sec_spirals}, we investigate in detail how spirals generate a mean-field dynamo in the presence of ambipolar diffusion.  We study in Section \ref{sec_res_boxsize} the dependence of this dynamo on numerical details, in particular the resolution and box size. Finally, in Section \ref{sec_conclusions}, we discuss the implications of our work for the evolution of protoplanetary and AGN discs.

\section{Model and numerical setup}
\label{sec_model}
\subsection{{Governing equations}}

We perform numerical simulations of gravito-turbulence and dynamo in a poorly ionised, differentially rotating disc using the local shearing-box approximation. In this model, a small patch of the disc, co-orbiting with angular velocity $\bb{\Omega}=\Omega_0\ez$ at a fixed radial location $R_0$, is represented by Cartesian coordinates $(x,y,z)$, with $x=R-R_0$ and $y=R_0\phi$ corresponding to the radial ($R$) and azimuthal ($\phi$) directions of a cylindrical coordinate system. Differential rotation is included through the Coriolis force and a background linear shear $\bb{v}_0 = -Sx\ey$, where $S \equiv -(\rmd\Omega/\rmd\ln R)_{R_0}$ is the local shear frequency; we use $S=(3/2)\Omega_0$, corresponding to Keplerian rotation. We adopt an ideal equation of state for the fluid in the disc, with the pressure $P$ and the mass density $\rho$ related by $P=\rho c^2_{\rm s}$, where $c_{\rm s} \equiv (\gamma P/\rho)^{1/2}$ is the sound speed and $\gamma$ is the ratio of specific heats (${=}5/3$). The internal energy $U\equiv P/(\gamma-1)$ of the disc is lost through radiative cooling at the rate $\Lambda^-$. An initially weak magnetic field $\bb{B}$ with zero net vertical flux is frozen into the minority charged species, which drift relative to the bulk neutral fluid at a rate controlled by collisional drag and parametrized through the field-strength-dependent ambipolar diffusivity $\eta_{\rm A}$.
%

Under these conditions, the equation governing the disc evolution are
\begin{gather}
    \pD{t}{\rho} + \grad\bcdot(\rho\bb{v}) = 0 , \label{mass_eq} \\*
    \pD{t}{(\rho\bb{v})} + \grad\bcdot(\rho\bb{v}\bb{v}) = - 2\bb{\Omega}\btimes\rho\bb{v} - \rho\grad\Phi - \grad P + \bb{J}\btimes\bb{B} , \label{ns_eq} \\*
    \pD{t}{\bb{B}} - \grad\btimes(\bb{v}\btimes\bb{B}) = \grad\btimes\bigl[ \eta_{\rm A} (\bb{J}\btimes\eb)\btimes\eb \bigr] , \label{magnetic_eq} \\*
    \pD{t}{U} + \grad\bcdot(U\bb{v}) = -P \grad\bcdot\bb{v} - \Lambda^- , \label{int_energy_eq}
\end{gather}
%
%
where $\bb{J}=\grad\btimes\bb{B}$ is the current density and $\eb\equiv\bb{B}/B$ is the magnetic-field unit vector. We decompose the fluid velocity $\bb{v}$ into its background and fluctuating parts:
\begin{equation}
    \bb{v} = \bb{v}_0 + \bb{u} = -\frac{3}{2} \Omega_0 x\ey + \bb{u} .
\end{equation}
%
The potential $\Phi$ in equation~\eqref{ns_eq} is the sum of the tidal potential in the local frame,
\begin{equation}
    \Phi_{\rm c} = \frac{1}{2}\Omega^2_0 z^2-\frac{3}{2}\Omega^2_0 x^2 , \label{eqn:phic}
\end{equation} 
and the gravitational potential of the disc itself, $\Phi_{\rm s}$. The latter obeys the Poisson equation 
\begin{equation}\label{poisson_eq}
    \mathbf{\nabla}^2\Phi_{\rm s} = 4\pi G\rho \,;
\end{equation}
its dynamical importance is quantified by the Toomre parameter
\begin{equation}\label{toomre}
Q \equiv \dfrac{c_s\Omega_0}{\pi G \Sigma} ,
\end{equation} 
where $\Sigma = \int\rmd z \, \rho$ the surface mass density of the disc.


For most of our simulations we assume a dependence of the radiative cooling on the temperature $T$ that is proportional to $T^4$:
\begin{equation}\label{eq_T4cool}
    \Lambda^-=\dfrac{U}{\tau_0} \left(\dfrac{T}{T_0} \right)^3 ,
\end{equation}
where $T_0 \propto c_{s_0}^2$ is the initial and uniform background temperature of the disc and $\tau_0$ is a uniform and constant coefficient that represents the effective cooling timescale when $T=T_0$. This prescription corresponds to an extreme case since cooling is not expected to be steeper than $T^4$. The steepness of this cooling law ensures that a thermal equilibrium between explicit heating and cooling always exists, independently of the location of the vertical boundary. By contrast, most numerical studies of gravito-turbulence in the existing literature have instead adopted the linear cooling function
\begin{equation}\label{eq_lincool}
    \Lambda^-=\dfrac{U}{\tau_c} ,
\end{equation}
with the characteristic timescale $\tau_c$ referred to as the `cooling time'. To make comparison with those studies, and to show that our qualitative results are insensitive to the choice of cooling law, we also performed a few simulations using equation~\eqref{eq_lincool}. Explicit thermal conduction is neglected.

Finally, we quantify the ambipolar diffusivity $\eta_A$ in the induction equation (\ref{magnetic_eq}) using the dimensionless ambipolar Els\"{a}sser number
\begin{equation}
\label{eq_Am}
    {\rm Am} \equiv \frac{v^2_{\rm A}}{\Omega_0\eta_{\rm A}} ,
\end{equation}
where $v_{\rm A} \equiv B/\sqrt{\rho}$ is the Alfv\'{e}n speed. When the only charged species in the disc are the ions and the electrons, $\eta_{\rm A} \approx v^2_{\rm A} \tau_{\rm ni}$, where $\tau_{\rm ni}$ is the neutral-ion collision timescale (proportional to the inverse of the ion mass density). In this case, ${\rm Am} \approx (\Omega_0 \tau_{\rm ni})^{-1}$ quantifies the statistical number of times that a neutral particle collides with an ion per orbital timescale; large values of ${\rm Am}$ indicate that the neutrals are well coupled to the ions (and thus to the magnetic field). Under more general conditions, $\eta_{\rm A}$ is obtained by solving a multi-species chemical network in which various ionisation sources (X-rays, cosmic rays, radioactive decay, FUV from the protostar) compete with charge recombination occurring both in the gas phase and on grain surfaces. For this first study of gravito-turbulence and dynamo in poorly ionised discs, we have opted for simplicity by crudely setting ${\rm Am}= \const$, effectively assuming a constant density for the charged species. 
In appendix \ref{appendixC}, we adopt an alternative prescription with ${\rm Am}=\const$ in the disc midplane and ${\rm Am}=\infty$ above a specified height, to mimic the presence of a well-ionised surface layer. The resulting simulation demonstrates that our results are similar whether or not such an ionised layer is present.

\subsection{Numerical methods}

We solve equations \eqref{mass_eq}--\eqref{int_energy_eq} in a three-dimensional (3D) shearing-periodic domain of size $(L_x,L_y,L_z)$, discretised on a mesh of $(N_X,N_Y,N_Z)$ grid points. To increase the trustworthiness of our results, we have used two different and independent codes, {\tt PLUTO} \citep{mignone2007} and  {\tt Athena++} \citep{stone20}; we perform several comparison runs which are detailed in Section \ref{sec_saturation}. Both codes use conservative finite-volume methods that solve the approximate Riemann problem at each inter-cell boundary and are well adapted to highly compressive flows (including shocks). The Riemann problem is handled by the HLLD solver, which is suitable for MHD. An orbital advection algorithm (similar to that detailed in \citet{migone12}) is used to increase the computational speed and reduce numerical dissipation.  The divergence of $\mathbf{B}$ is forced to 0 by a constrained-transport algorithm. In most of our simulations, the electromotive force (EMF) at the edge of each cell is reconstructed by the {\tt UCT\_CONTACT} averaging procedure (in both {\tt PLUTO} and {\tt Athena++}). Details about this method can be found in the online {\tt PLUTO} documentation and in \citet{gardiner05}. Both codes conserve the total energy and so the heat equation is not solved directly as in equation (\ref{int_energy_eq}).
{Heating by compression, magnetic diffusion, and shocks are all captured implicitly through the conservation of total energy, while cooling is treated as an additional sink term.}
Finally, the temporal integration is handled by a second -order-accurate Runge--Kutta method in {\tt PLUTO}, while it is handled by a second-order-accurate van Leer predictor-corrector method in {\tt Athena++}.

\subsubsection{Implementation of self-gravity in {\tt PLUTO}}

To compute the self-gravitational potential $\Phi_{\rm s}$ in {\tt PLUTO}, we take advantage of the shear-periodic boundary conditions. Following \citet{riols16a} and \citet{riols17b}, we first shift back the density in $y$ to the time at which it was last periodic. Then, for each plane at altitude $z_k$, we compute the direct 2D Fourier transform of the density $\widehat{\rho}_{k_x,k_y} (z)$, where $k_x$ and $k_y$ are the $z$-dependent radial and azimuthal wavenumbers. Using equation~\eqref{poisson_eq}, it is straightforward to show that the Fourier coefficients of the gravitational potential satisfy the Helmholtz equation
\begin{equation}
\left( \DD{z}{} - k^2 \right) \widehat{\Phi}_{k_x,k_y} (z) = 4\pi G \widehat{\rho}_{k_x,k_y} (z) 
\label{eq_helmholtz}
\end{equation} 
with $k^2 = k_x^2 + k_y^2$. This differential equation is solved in the complex plane by means of a fourth-order finite-difference scheme and a direct inversion method. The discretized system takes the form of a linear problem $\mathsf{AX}=\mathsf{B}$, where $\mathsf{X}$ is a vector representing the $z$-profile of the discretized potential, $\mathsf{A}$ is a penta-diagonal matrix, and $\mathsf{B}$ is a column vector containing the right hand side of equation \eqref{eq_helmholtz} and extra coefficients setting the boundary conditions. We use a fast algorithm involving $O(N_Z)$ flops to invert the matrix and obtain the discretized coefficients $\widehat{\Phi}_{k_x,k_y} (z_k)$. For each altitude $z_k$, we finally compute the inverse Fourier transform of the potential and shift it back to the initial sheared frame. 

The gravitational acceleration $-\rho\grad\Phi_{\rm s}$ is then obtained by computing the derivative of the potential in each direction using a fourth-order, cell-centred, finite-difference method. The result is implemented as a source term in the discretized version of the momentum equation \eqref{ns_eq}.

This implementation was tested by \citet[][see their Appendices A and B]{riols17b} using a stratified disc equilibrium and its linear stability.

\subsubsection{Implementation of self-gravity in {\tt Athena++}}


{
In {\tt Athena++}, we solve for $\Phi_{\rm s}$ using a discrete form of the Poisson equation in which the derivatives in the Laplace operator are computed with second-order finite differences. For example, the (continuous) second derivative $\partial_x^2 \Phi_{\rm s}$ is replaced with the discrete operator $\widehat\partial_x^2 \Phi_{{\rm s},i} \equiv (\Phi_{{\rm s},i+1}-2\Phi_{{\rm s},i}+\Phi_{{\rm s},i-1})/(\Delta x)^2$, where $i$ is the cell index in the $x$ direction and $\Delta x$ is the grid spacing. $\partial_x^2$ and $\widehat\partial_x^2$ share the same eigenfunctions $\exp({\rm i}k_x x)$, but the eigenvalue for $\widehat\partial_x^2$ is $-\widehat{k}_x^2 \equiv 2\left[\cos(k_x \Delta x)-1\right]/(\Delta x)^2$ instead of $-k_x^2$. (Their difference vanishes at long wavelengths, $|k_x\Delta x|\to 0$.)
Similar to the {\tt PLUTO} implementation, to compute $\Phi_{\rm s}$ we do a shifted 2D Fourier transform in $x,y$ and solve the resulting Helmholtz equation in $z$, where $k^2$ in equation \eqref{eq_helmholtz} is replaced by $\widehat{k}_x^2+\widehat{k}_y^2$ and ${\rm d}^2/{\rm d}z^2$ is computed with a second-order finite difference, to be consistent with the treatment in $x,y$ plane.
}

Instead of calculating $-\rho\grad\Phi_{\rm s}$ at cell centres and directly applying it as a source term, {\tt Athena++} calculates the momentum flux due to self gravity using the (second-order accurate)
gravitational stress evaluated at cell interfaces. This treatment better conserves momentum, and is more accurate than applying the gravitational acceleration as a source term whenever the density is
discontinuous or under-resolved.

We have tested this implementation of self gravity in {\tt Athena++} using various stratified disk equilibria and their linear stability (see appendix \ref{appendixA}).

\subsection{Numerical setup and free parameters}

In our codes, the unit of time is $\Omega^{-1}$ the orbital frequency and the unit of length is $H_0  \equiv c_{\rm s0}/\Omega_0$ the standard disc scale height (corresponding to the actual scale height of an isothermal disc without self-gravity; the true scale height with self-gravity is smaller than this value).
For most of our simulations, we use a box of size $L_x = L_y= 20 H_0$ and $L_z=9 H_0$. The horizontal size is chosen so that three or four spiral structures may be captured in the box at the same time, {with the caveat that the shearing-box approximation, which assumes $L_x,L_y\ll$ disc radius, is only marginally justified at such large box sizes}. The vertical size is chosen so that the energy that escapes through the vertical boundary is not too large in comparison to the energy lost through radiative cooling. Our fiducial numerical resolution is $256 \times 256 \times 144$ cells, which corresponds to 13 cells per $H_0$ in the horizontal directions and 16 cells per $H_0$ in the vertical direction. (This horizontal box size and resolution are identical to those of \citetalias{riols19b}.) In Section \ref{sec_res_boxsize} we study the dependence of our results on resolution and box size.

The boundary conditions are periodic in $y$ and shear-periodic in $x$. In the vertical direction, for hydrodynamic variables (velocity,  density, and pressure), we use a simple outflow boundary condition and suppress inflow by setting $v_z=0$ in the ghost zones if the local velocity points towards the disc. For the magnetic field we use a vertical boundary condition in which the field is `mirrored' into the ghost zones with $B_x$ and $B_y$ changing sign (such that $B_x=B_y=0$ on the boundary). We checked that using the mirror symmetry or forcing $B_x=B_y=0$ in all ghost zones gives the same results.  For the self-gravitating potential, {\tt PLUTO} assumes that 
\begin{equation}
\D{z}{} \widehat{\Phi}_{k_x,k_y} (\pm {L_z}/{2}) = \mp k \widehat{\Phi}_{k_x,k_y} (\pm {L_z}/{2}) .
\end{equation}
This condition is an approximation of the Poisson equation in the limit of low density and ensures that the potential does not diverge at infinity. {\tt Athena++} simply sets $\widehat{\Phi}_{k_x,k_y}=0$ in the ghost zones for $k_x\neq0$ and $k_y\neq0$. This boundary condition for the potential  matches that of {\tt PLUTO} at $z\rightarrow \pm \infty$ (if the box had infinite size in $z$).

To compensate for the loss of mass via outflow, the gas is replenished near the midplane so that the total mass in the box (and therefore $\Sigma$) is maintained constant during the time of each simulation.  When adding mass, we change the density of all cells by a constant factor, while fixing momenta and total energy density. 

To avoid excessively small timesteps, especially when a magnetic field is present, a finite density floor  $\rho_{\rm floor}=10^{-4}$ is used. Such a density floor can be problematic in a relatively tall box, because the gravitational acceleration (accounting for both self gravity and the vertical component of stellar gravity) acting on the floor density creates unphysical acceleration that continually pulls material towards the midplane and generates artificial heating. We alleviate this problem by calculating the gravitational force as $-(\rho-\rho_{\rm floor})\grad\Phi$ instead of $-\rho\grad\Phi$. Large values of magnetic diffusivity can also cause prohibitive timesteps, and so we have capped $\eta_A$ so that
\begin{equation}
    \eta_A= \text{min}\left\lbrace \eta_{\text{cap}}, \dfrac{1}{\text{Am}}\dfrac{B^2}{\rho \Omega}\right\rbrace ,
\end{equation}
where $\eta_{\text{cap}}=3$ (in units of $c_{s_0}^2/\Omega$) for $\text{Am}=1$ and $\eta_{\text{cap}}=1$ for larger $\text{Am}$. For the {\tt Athena++} simulations that employ double resolution $(512 \times 512 \times 288)$, ambipolar diffusion is calculated with a super-time-stepping algorithm \citep{alexiades96} to reduce cost. We have tested at our fiducial resolution  that super-time-stepping does not affect the results.

Our initial conditions depend on whether or not a magnetic field is present. For our purely hydrodynamical simulations, we start from a self-gravitating disc equilibrium in which the strength of self gravity is specified by $Q_0 = 1$. The equilibrium is initially isothermal with temperature $T_0 = c_{s_0}^2/\gamma$ (in code units). The density in the midplane is fixed to 1 so that the surface density is $\Sigma=0.917$. Random, non-axisymmetric, finite-amplitude density and velocity perturbations are injected to trigger a gravito-turbulent state. For our MHD runs, we use a gravito-turbulent steady state taken from a hydrodynamical simulation as the initial condition and add in a weak seed toroidal field. In this case, the seed field is of the form
\begin{equation}
\label{eq_Byinit}
B_y = B_{y_0} \sin (2\pi z/L_z)
\end{equation}
and its strength is characterized by the initial midplane 
\begin{equation}
\beta_{y_0} = \dfrac{2P_0}{B_{y_0}^2}=\dfrac{2\rho_0 c_{s_0}^2}{\gamma B_{y_0}^2} .
\end{equation}
The list of runs is shown in Table \ref{table1} with the different input parameters specified in the columns.

\subsection{Diagnostics}

\subsubsection{Averaged quantities}
\label{diagnostics_av}

To analyse the statistical behaviour of the turbulent flow, we define the standard and weighted box averages
\begin{equation}
\langle X \rangle =\frac{1}{V}\int_V \rmd V \, X  \quad \text{and} \quad \langle X \rangle_W = \frac{1}{V\langle\rho\rangle} \int_V\rmd V\, \rho  X ,
\end{equation}
respectively, where $V = L_x L_y L_z$ is the volume of the box. For example, $E_k=\frac{1}{2}\langle\rho {u}^2\rangle$ and $E_m=\frac{1}{2}\langle{B}^2\rangle$ are the box-averaged kinetic and magnetic energies, respectively. We also define the horizontally averaged vertical profile of a variable:
\begin{equation}
\overline{X}(z) = \frac{1}{L_xL_y} \iint\rmd x\rmd y \, X .
\end{equation}
This averaging will be useful when discussing mean-field equations in the context of dynamo (see \S\ref{sec_MF}). 

There are three important box-averaged quantities with which we characterise the gravito-turbulent steady state. First, we define the 2D Toomre parameter
\begin{equation}
\label{eq_toomre}
Q_W \equiv \dfrac{\langle c_s \rangle_W\Omega_0}{\pi G \Sigma },
\end{equation}
where $\Sigma=L_z \langle\rho\rangle$ is the average surface density of the disc. This quantity is directly relevant for characterising the stability of GI spiral modes.  Second, we follow the custom of quantifying the efficiency of the angular-momentum transport in terms of an $\alpha$ parameter, although one which combines the contributions from the Reynolds ($H_{xy}$), gravitational ($G_{xy}$), and Maxwell ($M_{xy}$) stresses:
\begin{align}\label{def_alpha}
\alpha\equiv \dfrac{\langle
  H_{xy}+G_{xy}+M_{xy} \rangle}{\langle P\rangle},
\end{align} 
where
\[
H_{xy}=\rho u_xu_y,
\quad G_{xy}=\frac{1}{4\pi G} \pD{x}{\Phi} \pD{y}{\Phi}, \quad{\rm and}\quad M_{xy}=-B_xB_y.
\]
Finally, the average disc thermodynamics is described by the average effective cooling {timescale}
\begin{equation}
\label{eq_taueff1}
    \tau_{\rm eff} \equiv \dfrac{\langle U \rangle}{\langle \Lambda^- \rangle} .
\end{equation}
For simulations with a linear cooling law, $\tau_{\rm eff}=\tau_c$, while for simulations with $T^4$ cooling (equation \ref{eq_T4cool}),
\begin{equation}\label{eq_taueff}
\tau_{\rm eff}= \tau_0  {\langle U \rangle}{\left\langle U\left(\dfrac{T}{T_0} \right)^3\right \rangle}^{-1}.
\end{equation}
This last relation implies that the average effective cooling time can be different from $\tau_0$ depending on how the temperature deviates from the isothermal condition $T=T_0$.  This is important in the context of making direct comparisons between runs with $T^4$ cooling and runs with linear cooling. To facilitate such comparisons, we first compute the turbulent steady state using $T^4$ cooling, measure $\tau_{\rm eff}$ via equation \eqref{eq_taueff}, and then set $\tau_{\rm c}=\tau_{\rm eff}$ in a complementary run using the linear cooling law \eqref{eq_lincool}.


\subsubsection{Fourier decomposition and spectra}

To analyse the structure of the flow and, in particular, the role of spiral waves and small-scale dynamics in the dynamo process, we employ a 2D Fourier decomposition of the fields. Namely, we denote by  $\widehat{\bb{u}}^e(k_x,k_y,z)$ and  $\widehat{\bb{B}}^e(k_x,k_y,z)$ the horizontal 2D decomposition (in
Eulerian wavenumbers $k_x$ and $k_y$) of the turbulent velocity and magnetic field, for a given altitude $z$, and averaged over a given period of time. In this mathematical representation, the fields are a sum of a {mean} ($k_x=k_y=0$) mode, axisymmetric modes with $k_y=0$ ($k_x\neq0$), and non-axisymmetric modes with $k_y\neq0$ (commonly referred as `shearing
waves'). These Fourier decompositions are calculated via a method described by \citet[][\S 2.5.2]{riols17b} using the FFT algorithm. The 2D kinetic and magnetic energy power spectra are then defined as 
\begin{equation}
E_{K}(k_x,k_y,z)=\frac{1}{2} \Sigma H_0^{-1} \Bigl|\widehat{\bb{u}}^e(k_x,k_y,z)\Bigr|^2,
\end{equation}
\begin{equation}
E_{M}(k_x,k_y,z)=\frac{1}{2} \Bigl|\widehat{\bb{B}}^e(k_x,k_y,z)\Bigr|^2 .
\end{equation}

\subsubsection{Phase-folding technique}\label{phase_folding}

As shown by \citet{riols18}, when 3D dynamics are allowed in simulations of gravito-turbulence, the customary spiral density waves are accompanied by characteristic hydrodynamical motions such as vertical rolls that play an important role in amplifying magnetic fields. So far the analysis of these  motions, as well as of the geometry of the magnetic around the spiral arms, has focused on single snapshots of the flow and field (e.g., see Fig.~15 of \citetalias{riols19b}). However, the strong gravito-turbulence makes it difficult to interpret these shapshots. To filter out these turbulent fluctuations and see more clearly how spiral waves (and their accompanying vertical rolls) shape the magnetic field, we phase-fold our data as follows. First, we take a series of snapshots from moments when the fields are exactly periodic in $x$ (typically once every $2\Omega^{-1}$). From a Fourier decomposition of the 2D surface density, we then identify the most prominent spiral mode over the course of the simulation, say, $k_y=2\pi/L_y$ and $k_{x}=k_{x_s}$. For each snapshot we calculate the phase of this mode and shift all the fields in $x$ to ensure that this phase is reduced to zero (in that way we align the phase of the spiral waves). We then stack the shifted fields upon one another to obtain a time average. We perform an additional average of the data over the azimuthal ($y$) direction along the spiral structure, i.e., along $(k_{x_s}/k_{x_0}) x + y = \const$ (this is done by shifting the data in $x$ and averaging in $y$). The result is a phase-folded poloidal map ($x$,$z$) of the flow, that we will note with a superscript $s$,  where systematic perturbations associated with the most prominent spiral mode are preserved but turbulent fluctuations uncorrelated with the spiral mode are suppressed by the averaging procedure. 

\subsubsection{Mean-field equations and EMFs}\label{sec_MF}

One of our principal goals is to characterise the ability of gravito-turbulent motions to supply and sustain a mean field dynamo. Therefore, a key quantity is the mean magnetic field  $\overline{\bb{B}}(z)$, averaged in the $x$ and $y$ directions, corresponding to the $k_x=k_y=0$ Fourier component of $\bb{B}$. The equations governing its radial ($x$) and toroidal ($y$) projections are: 
\begin{equation}\label{eq_Bxmean}
\pD{t}{\overline{B}_x} = -\pD{z}{} \left( \overline{\mc{E}}_y + \overline{\mc{E}}_y^{\rm NI} \right) ,
\end{equation}
\begin{equation}\label{eq_Bymean}
\pD{t}{\overline{B}_y} = -S\overline{B}_x + \pD{z}{} \left( \overline{\mc{E}}_x + \overline{\mc{E}}_x^{\rm NI} \right) ,
\end{equation}
where $\overline{\bb{\mc{E}}}(z)= \overline{\bb{u}\btimes\bb{B}}$ is ideal component of the horizontally-averaged electromotive force and  
\begin{equation} 
\overline{\bb{\mc{E}}}^{\rm NI}(z) = \overline{\eta_{\rm A} (\bb{J} \btimes\eb \btimes \eb)}
\end{equation}
its non-ideal component. The ideal EMF can be decomposed into a sum of a `spiral' component $\overline{\bb{\mc{E}}}^\star = \overline{\bb{u}^\star\btimes\bb{B}^\star}$, where the `$\star$' superscript  refers to the phase-folded profile of the fields (see Section \ref{phase_folding}) and a turbulent component defined by $\overline{\bb{\mc{E}}}^t= \overline{\bb{u}\btimes\bb{B}} - \overline{\bb{u}^\star\btimes\bb{B}^\star}$. 

We remind the reader that the mean radial and toroidal fields in the box, respectively $\langle B_x \rangle$ and $\langle B_y \rangle$, equal to the $z$-average of $\overline{{B}}_{x}(z)$ and  $\overline{{B}}_{y}(z)$, are free to evolve because of the open boundaries at the top and bottom of the computational domain.

\section{Properties of the gravito-turbulent dynamo}\label{sec_saturation}

\subsection{Initial flows: hydrodynamical simulations}\label{hydro_sims}

A customary way to study the dynamo process in simulations is to compute the (unmagnetized) hydrodynamic state of fully developed turbulence, introduce a zero-net flux magnetic seed field, and see whether that field grows or decays over time. Following this scheme, we first simulate purely hydrodynamical gravito-turbulence, with the aim of comparing two different cooling laws (linear and $T^4$) as well as both codes, {\tt PLUTO} and {\tt Athena++} (a similar code comparison was done by \citet{riols17b} using {\tt PLUTO} and {\tt RODEO}). 
The key parameters that control the strength of the turbulence are the cooling times $\tau_0$ (when $T^4$ cooling is used; see eqn~\ref{eq_T4cool}) and $\tau_c$ (when linear cooling is used; see eqn~\ref{eq_lincool}). For $T^4$ cooling, we use the fiducial $\tau_0=10\Omega^{-1}$ (runs AT-GHydro and PL-GHydro). For linear cooling, we choose $\tau_c=11.5\Omega^{-1}$, which matches the $\tau_{\rm eff}$ measured in the $T^4$ runs (runs AT-GHydro-lin and PL-GHydro-lin, see \S\ref{diagnostics_av} and Table \ref{table2}).

Table \ref{table2} summarises some box-averaged quantities obtained with {\tt PLUTO} and {\tt Athena++}. We find excellent agreement between the two codes for the hydrodynamical runs, despite the different methodology used to compute self-gravity. In particular, in both cases the Toomre parameter settles at $Q_W = 1.12$ and the ratio of the gravitational and Reynolds stresses is around 2.5. The box-averaged quantities do not seem to depend too much on the cooling law, although the kinetic energy and Reynolds stress are slightly larger in the linear cooling runs. 

From energy conservation and the fact that outflows are observed to be very weak in the purely hydrodynamical case, one may expect the  time-averaged stress to follow the \citet{Gammie2001} relation,
\begin{equation}
\label{gammie_eq}
\alpha\simeq\dfrac{1}{q\Omega (\gamma-1) \tau_{\rm eff}} = \dfrac{1}{\Omega \tau_{\rm eff}}  \quad (\gamma=5/3, \, q=3/2) 
\end{equation} 
The values of $\alpha$ are shown in Table \ref{table2}; they follow the \citet{Gammie2001} relation within a relative error of less than $5 \%$.

%
%
\begin{figure}
    \centering
    \includegraphics[width=.5\textwidth]{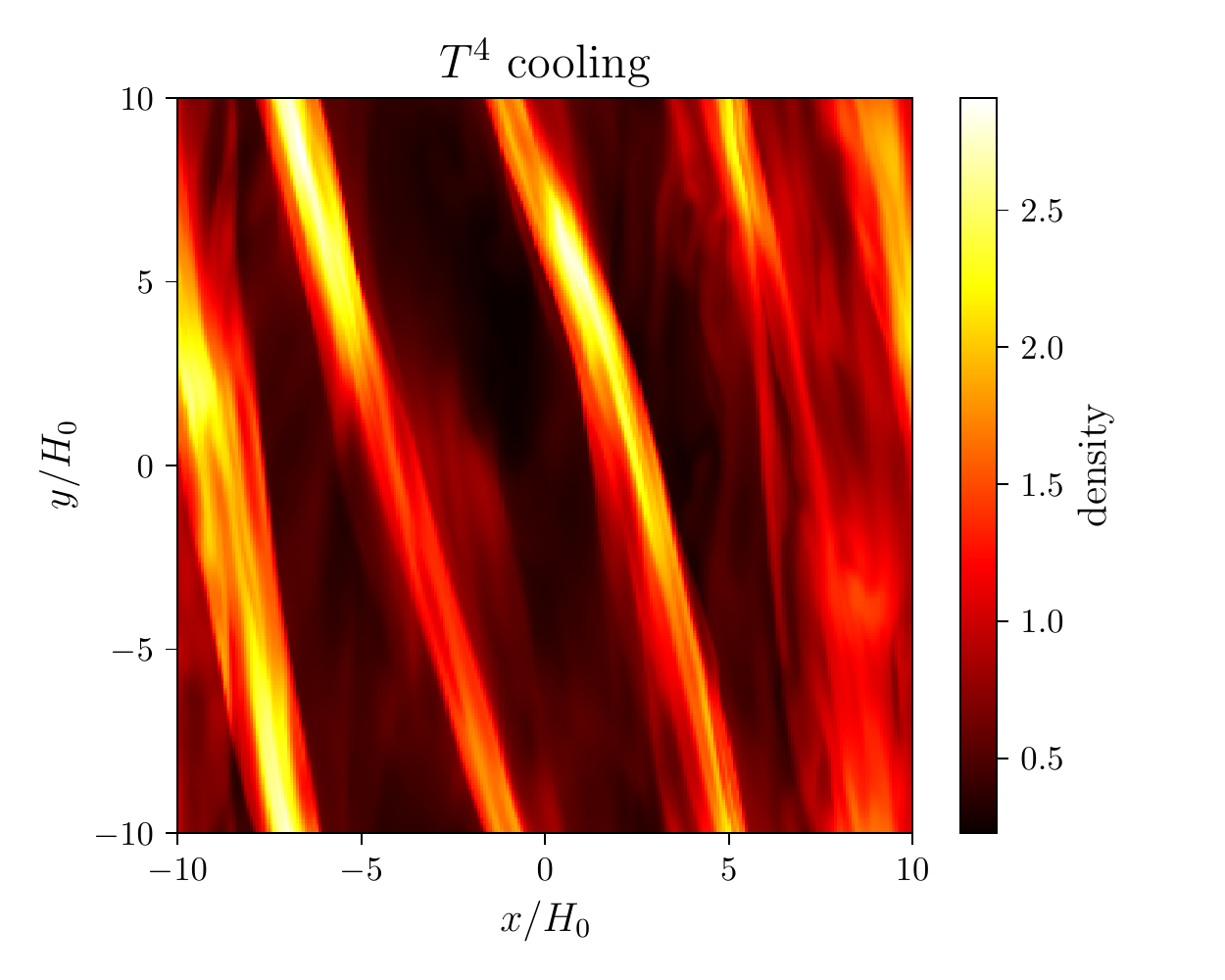}
    \includegraphics[width=.5\textwidth]{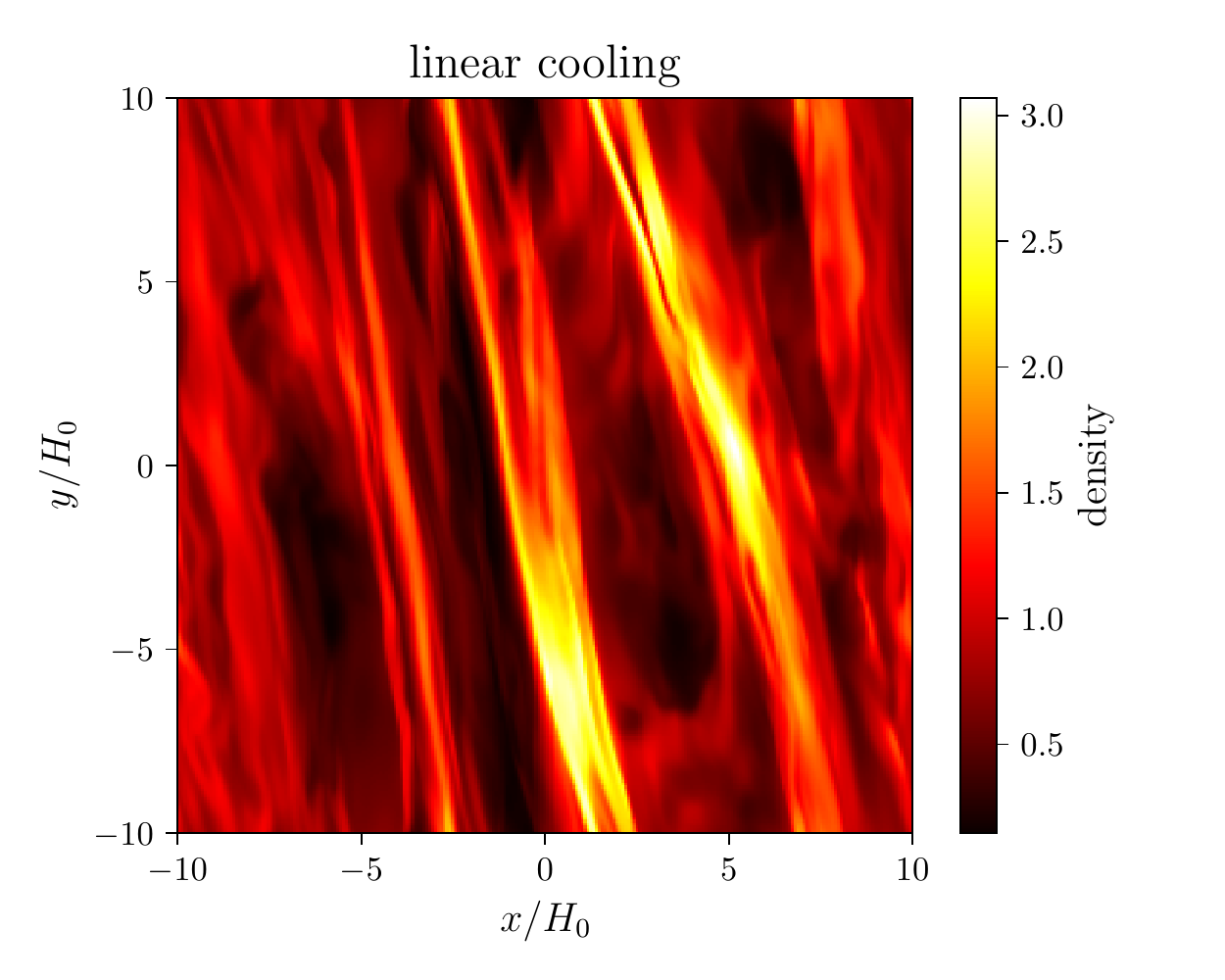}
    \caption{Top: Snapshot of midplane density for AT-GHydro ($T^4$ cooling) at an arbitrary time in steady state. Density perturbation is dominated by GI-driven spirals density waves. Bottom: same but for run AT-GHydro-lin (linear cooling). Small-scale fluctuations seem more visible in the latter case.}
    \label{fig_hyd}
\end{figure}

%
%
\begin{figure}
    \centering
    \includegraphics[width=.47\textwidth]{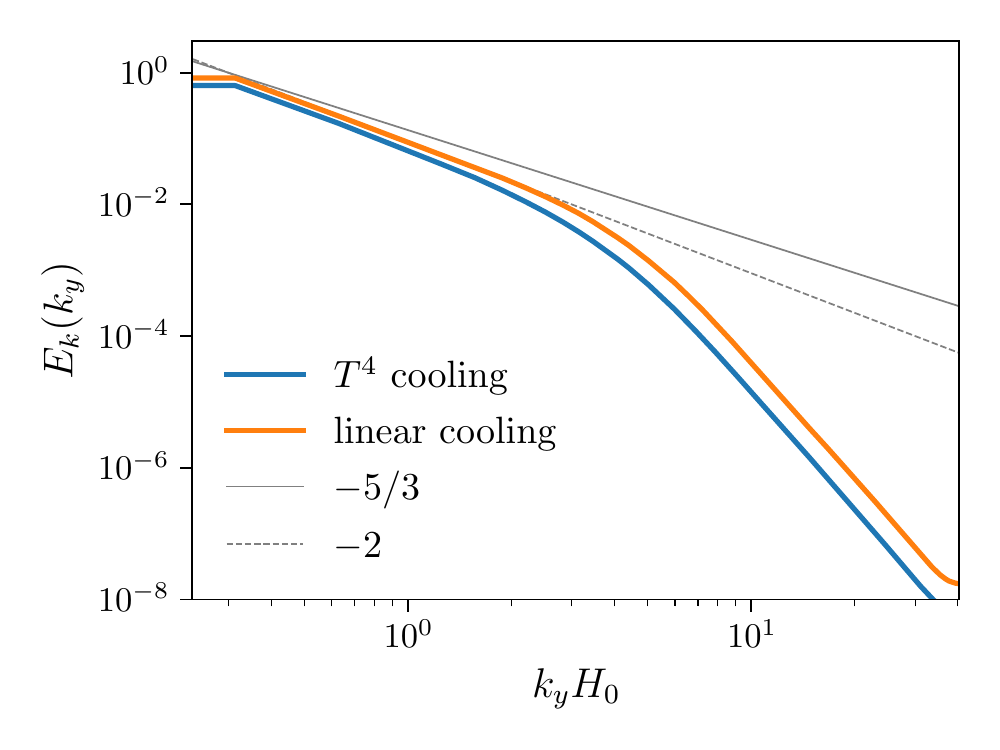}
    \caption{1D kinetic energy spectra $E_K$, averaged in time, $z$, and $k_x$ for AT-GHydro and AT-GHydro-lin. The two are very similar, except slightly higher amplitudes at small scale (high ${k_y}$) for the linear cooling run.}
    \label{fig_hyd_spectra}
\end{figure}

For both the linear- and $T^4$-cooling runs, the turbulence is supersonic, highly compressible, and characterised by large-scale spiral density waves, which are particularly strong for the values of $\tau_0$ and $\tau_{\rm c}$ considered. A snapshot of the density structure is shown in Fig.~\ref{fig_hyd} for the runs AT-GHydro and AT-GHydro-lin. We clearly identify three or four spiral waves with a dominant $k_y=k_{y_0}=2\pi/L_y$ mode. The shape of these spiral structures and their amplitude (or contrast) do not seem to depend significantly on the cooling law. Fig.~\ref{fig_hyd_spectra} shows the kinetic-energy spectra $E_{K}(k_y)$ averaged in time, $z$, and $k_x$ for the linear- and $T^4$-cooling runs. We find that the two spectra are very similar, except perhaps that small-scale motions are more prominent in the linear-cooling run. On large scales the power spectrum agrees closely with {$k^{-2}$}
but for smaller scales ($k_y H_0 \gtrsim 4$) the kinetic energy is steeper. These slopes are similar to those found by \citet{booth19}.  In both cases, spiral waves are accompanied by strong axisymmetric structures with $k_y=0$ and $k_{x}=k_{x_0}=2\pi/L_x$, which have been shown to trigger small-scale inertial modes via a parametric instability in GI turbulence \citep{riols17b}.

\begin{figure*}
\centering
\includegraphics[width=\textwidth]{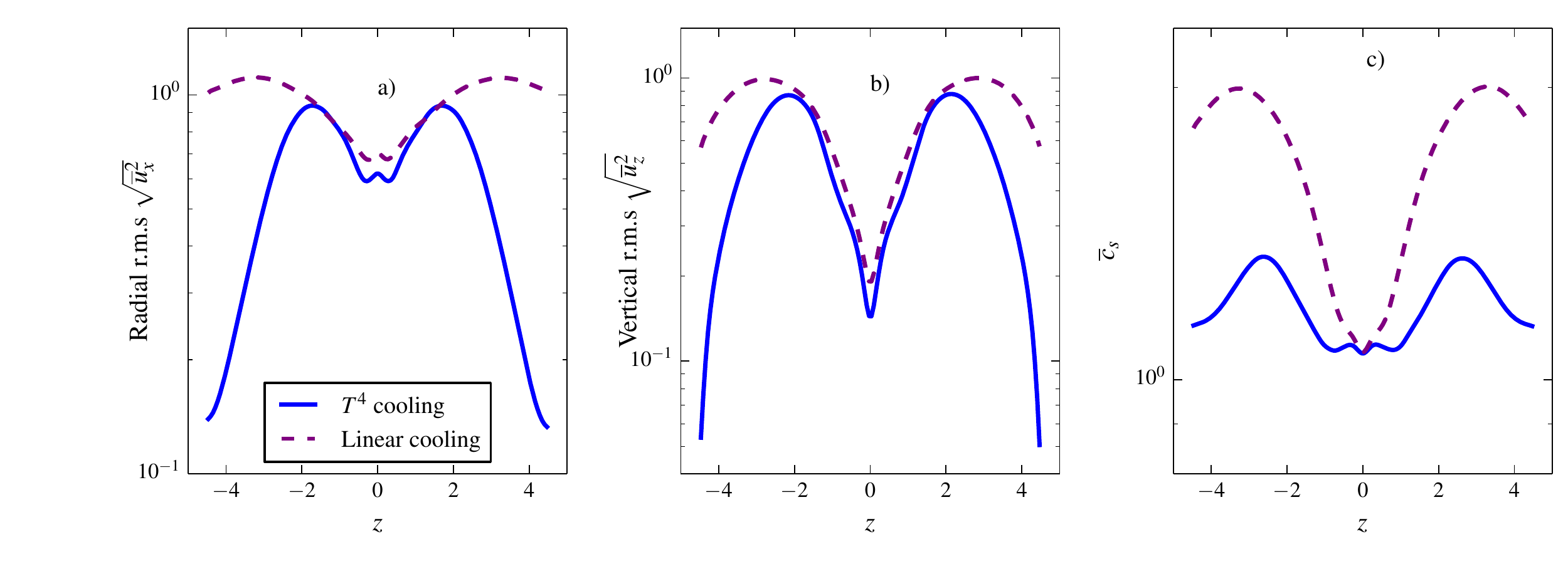}
 \caption{Vertical profile of r.m.s radial (left), vertical (centre) velocities  and sound speed (right). Quantities are averaged in $x$, $y$ and time from the {\tt PLUTO} hydrodynamical runs.}
\label{fig_rms}
 \end{figure*}

The left and central panels of Fig.~\ref{fig_rms} shows the vertical profiles of the root-mean-square (r.m.s.) radial and vertical velocities, $\sqrt{\overline{u}_x^2}$ and $\sqrt{\overline{u}_z^2}$ respectively, for run PL-GHydro ($T^4$ cooling; solid line) and run PL-GHydro-lin (linear cooling; dashed line). Around the midplane and below $z=2.5 H_0$, the two runs share the same behaviour. In the midplane ($z=0$), the velocity fluctuations are dominated by radial motions. However, higher up in the disc corona the vertical velocity component becomes comparable. These profiles issue from the combination of the vertical roll motions that accompany the spiral waves at $z\lesssim H_0$ \citep[see][]{riols18},
and small-scale inertial modes that attack these spirals at all
altitudes (but with some predominance at $z \gtrsim H_0$). The main difference between the two runs occurs near the box boundaries at $z \gtrsim 2.5H_0$. In the case of $T^4$ cooling, the vertical r.m.s.~velocity fluctuation is ${\sim}0.05c_{s_0}$, while for the case of linear cooling it is ${\sim}10$ times higher. As a result, the linear-cooling run exhibits significantly more outflow (although it remains small overall), with mass-loss rates {$\dot m/(\Sigma\Omega)$}${\approx}1.1 \times 10^{-4}$ (see Table \ref{table2}). We checked that this difference is obtained both with {\tt PLUTO} and with {\tt Athena++}. Finally, the right panel of Fig.~\ref{fig_rms} shows that the thermal structure of the disc is different depending on the cooling prescription. The corona is almost two times hotter in the case of a linear cooling. However, these differences are only minor, since the results (in particular the dynamo behaviour) depend only weakly on the cooling law (see \S\ref{cooling_law_dep}). 

%
%
\begin{center}
\begin{table*}
\centering 
\begin{tabular}{c c c c c c c c}
\hline
Run & Resolution & \makecell{$L_z$ \\ (in $H_0$)} & \makecell{Duration \\ (in $\Omega^{-1}$)} & \makecell{Transient \\ (in $\Omega^{-1}$)} & Cooling &  \makecell{$\tau_0$ or $\tau_c$ \\ (in $\Omega^{-1}$)} & Am \\
\hline  
	AT-GHydro 	& $256\times 256\times 144$ & 9 & 500 & 100 & $T^4$ & 10 & -- \\
    AT-GHydro-lin &  $256\times 256\times 144$ & 9 & 500 &  100 & Linear & 11.5 & -- \\    
    AT-ZNF-ideal & $256\times 256\times 144$ & 9 & 2000 & 500 & $T^4$ & 10 & -- \\
    AT-ZNF-ideal-lin & $256\times 256\times 144$ & 9 & 1000 & 450 & Linear & 9.5 & -- \\
    AT-ZNF-ideal-HR & $512\times 512\times 288$ & 9 & 400 & 100 & $T^4$ & 10 & -- \\
    AT-ZNF-Am100 & $256\times 256\times 144$ & 9 & 500 & 200 & $T^4$ & 10 & 100 \\
    AT-ZNF-Am100-HR & $512\times 512\times 288$ & 9 & 700 & 0 & $T^4$ & 10 & 100 \\
    AT-ZNF-Am10 & $256\times 256\times 144$ & 9 & 450 & 200 & $T^4$ & 10 & 10 \\
    AT-ZNF-Am10-lin & $256\times 256\times 144$ & 9 & 100 & 20 & Linear & 7 & 10 \\
    AT-ZNF-Am10-HR & $512\times 512\times 288$ & 9 & 100 & 0 & $T^4$ & 10 & 10 \\
    AT-ZNF-Am1 & $256\times 256\times 144$ & 9 & 500 & 200 & $T^4$ & 10 & 1 \\
    \hline
    PL-GHydro 	& $256\times 256\times 144$ & 9 & 250 & 50 & $T^4$ & 10 & -- \\
    PL-GHydro-Lz12 	& $256\times 256\times 144$ & 12 & 470 & 50 & $T^4$ & 10 & -- \\
    PL-GHydro-lin &  $256\times 256\times 144$ & 9 & 400 &  0 & Linear & 11.5 & -- \\    
    PL-ZNF-ideal & $256\times 256\times 144$ & 9 & 800 & 250 & $T^4$ & 10 & -- \\
    PL-ZNF-ideal-lin & $256\times 256\times 144$ & 9 & 660 & 200 & Linear & 11.5 & -- \\
    PL-ZNF-Am200 & $256\times 256\times 144$ & 9 & 300 & 50 & $T^4$ & 10 & 200 \\
    PL-ZNF-Am100 & $256\times 256\times 144$ & 9 & 480 & 260 & $T^4$ & 10 & 100 \\
    PL-ZNF-Am100-Lz12 & $256\times 256\times 144$ & 12 & 400 & 175 & $T^4$ & 10 & 100 \\
    PL-ZNF-Am100-lin & $256\times 256\times 144$ & 9 & 350 & 0 & Linear & 4 & 100 \\
    PL-ZNF-Am100-HR & $512\times 512\times 288$ & 9 & 160 & 100 & $T^4$ & 10 & 100 \\
    PL-ZNF-Am30 & $256\times 256\times 144$ & 9 & 240 & 0 & $T^4$ & 10 & 30 \\
    PL-ZNF-Am10 & $256\times 256\times 144$ & 9 & 460 & 200 & $T^4$ & 10 & 10 \\
    PL-ZNF-Am10-Lz12 & $256\times 256\times 144$ & 12 & 200 & 20 & $T^4$ & 10 & 10 \\
    PL-ZNF-Am10-lin & $256\times 256\times 144$ & 9 & 100 & 0 & Linear & 7 & 10 \\
    PL-ZNF-Am3 & $256\times 256\times 144$ & 9 & 200 & 20 & $T^4$ & 10 & 3 \\
\\                              
\end{tabular}  
\vspace{0.5cm}
\caption{List of hydrodynamical and zero-net-flux MHD simulations, combining {\tt Athena++} (AT-) and {\tt PLUTO} (PL-) runs. The column `transient' indicates the time before the simulation reaches a quasi-steady turbulent state. The parameters $\tau_0$, $\tau_c$ and Am are defined respectively in \ref{eq_T4cool}, \ref{eq_lincool} and \ref{eq_Am}. }
\label{table1}
\end{table*}  
\end{center}

\begin{center}
\begin{table*}
\centering 
\begin{tabular}{c c c c c c c c c c c}
\hline
Run & $\tau_{\rm eff}$ & $Q_W$  & $\langle U \rangle$ & $E_k$/$\langle U \rangle$ & $E_m$/$\langle U \rangle$ & $\langle H_{xy} \rangle$/$\langle U \rangle$ & $\langle G_{xy}\rangle$/$\langle U \rangle$ & $\langle M_{xy}\rangle$/$\langle U \rangle$ & $\alpha$  & $\dot{m}_w$/$(\Sigma \Omega)$\\
\hline  
	AT-GHydro & 11.91 &1.116  & 0.1173 & 0.2067 & 0 & 0.01646 & 0.03848 & 0 & 0.08241 & $-6.67\times10^{-6}$ \\
    AT-GHydro-lin &  11.5 & 1.146  & 0.1257 & 0.2851 & 0 & 0.01860 & 0.0400 & 0 & 0.08790 & $1.14 \times10^{-4}$  \\    
    AT-ZNF-ideal & 9.38 & 1.175  & 0.1297 & 0.1838 & 0.07449 & 0.01822 & 0.03105 & 0.02212 & 0.10708 & $1.15 \times10^{-4}$ \\
    AT-ZNF-ideal-lin & 9.5  & 1.442 & 0.2034 & 0.1092 & 0.1535 & 0.01806 & 0.01161 & 0.05048 &  0.1202 & $1.01 \times 10^{-3}$ \\
    AT-ZNF-ideal-HR & 13.14 &  1.126 & 0.1184 & 0.1410 & 0.03550 & 0.01277 & 0.02471 & 0.01321 & 0.07603 & $6.01 \times 10^{-5}$ \\
    AT-ZNF-Am100 & 4.31 & 1.376 & 0.1767 & 0.1596 & 0.6007 & 0.04185 & 0.007985 & 0.1258 & 0.2634 & $2.11 \times 10^{-3}$  \\
    AT-ZNF-Am100-HR & 5.67 & 1.297 & 0.1574 & 0.1472 & 0.3358 & 0.03043 & 0.01328 & 0.08580 & 0.1943 & $9.17 \times 10^{-4}$  \\
    AT-ZNF-Am10 & 7.01 & 1.229 & 0.1418 & 0.1627 & 0.2854 & 0.02367 & 0.02842 & 0.04959 & 0.1525 & $5.13 \times 10^{-4}$\\
    AT-ZNF-Am10-lin & 7 & 1.230 & 0.1503 & 0.2305 & 0.2841 & 0.0250 & 0.03167 & 0.05415 & 0.1662 & $1.04 \times 10^{-3}$  \\
    AT-ZNF-Am10-HR & 6.98 & 1.235 & 0.1427 & 0.1636 & 0.2603 & 0.02493 & 0.02667 & 0.04889 & 0.1507 & $3.97 \times 10^{-4}$ \\
    AT-ZNF-Am1 & 10.89 & 1.130 & 0.1204 & 0.2537 & 0.02023 & 0.01696 & 0.03887 & 0.004507 & 0.0905 & $1.94 \times 10^{-5}$ \\
    \hline
    PL-GHydro 	& 11.41 & 1.119 & 0.1180 & 0.2417 & 0 & 0.01673 & 0.04171 & 0 & 0.0876 & $-1.7\times10^{-6}$  \\
    PL-GHydro-Lz12 & 10.62 & 1.13 & 0.09024 & 0.248 & 0 & 0.01944 & 0.04363 & 0 & 0.0946 & $-4.38\times10^{-8}$ \\ 
    PL-GHydro-lin & 11.5  & 1.144 & 0.1255 &  0.2996 & 0 & 0.0199 & 0.0409 & 0 & 0.0911 & $1.10\times10^{-4}$  \\    
    PL-ZNF-ideal & 10.56 & 1.153 & 0.1246 & 0.1832 & 0.0473 & 0.01679 & 0.03118 & 0.01576 & 0.0955 & $7.75 \times 10^{-5}$ \\
    PL-ZNF-ideal-lin & 11.5 & 1.170 & 0.1307 & 0.213 & 0.0372 & 0.01707 & 0.03053 & 0.01342 & 0.09153 & $1.15 \times 10^{-4}$ \\
    PL-ZNF-Am200 & 7.41 & 1.234 & 0.1425 & 0.1712 & 0.1504 & 0.02269 & 0.02494 & 0.04565 & 0.139 & $2.81 \times 10^{-4}$\\
    PL-ZNF-Am100 & 3.93 & 1.398 & 0.1821 & 0.1700 & 0.6149 & 0.04505 & 0.006020 & 0.1397 & 0.2861 & $1.97 \times 10^{-3}$ \\
    PL-ZNF-Am100-Lz12 & 3.78 & 1.398 & 0.1367 & 0.1777 & 0.6574 & 0.04502 & 0.0074 & 0.1379 & 0.2854 & $8.32 \times 10^{-4}$\\
    PL-ZNF-Am100-lin & 4 & 1.562 & 0.2195 & 0.1802 & 0.6095 & 0.04740 & 0.0045 & 0.1537 & 0.3083 & $3.57 \times 10^{-3}$ \\
    PL-ZNF-Am100-HR & 4.81 & 1.354 & 0.1707 & 0.1536 & 0.3877 & 0.0331 & 0.0118 & 0.1030 & 0.2218 & -- \\
    PL-ZNF-Am30 & 6.20 & 1.2849 & 0.1537 & 0.1188 & 0.6269 & 0.02848 & 0.00680 & 0.08688 & 0.1832 & $1.21 \times 10^{-3}$  \\
    PL-ZNF-Am10 & 7.21 & 1.224 & 0.1403 & 0.1601 & 0.2864 & 0.02307 & 0.02601 &0.04993 & 0.1485 &  $4.32 \times 10^{-4}$ \\
    PL-ZNF-Am10-Lz12 & 7.46 & 1.220 & 0.1045 & 0.1582 & 0.2845 & 0.02152 & 0.0249 & 0.04725 & 0.1405 & $3.17 \times 10^{-4}$\\
    PL-ZNF-Am10-lin & 7 & 1.222 & 0.1476 & 0.2075 & 0.2899 & 0.02456 & 0.02456 & 0.05657 & 0.1585 & -- \\
    PL-ZNF-Am3 & 9 & 1.180 & 0.1235 & 0.17652 & 0.07058 & 0.01727 & 0.03389 & 0.01472 & 0.1 & -- \\
\\                              
\end{tabular}  
\vspace{0.5cm}
\caption{Box-averaged and flux quantities measured in the simulations of Table \ref{table1}. $\tau_{\rm eff}$ is the effective cooling time defined in \ref{eq_taueff1}. $Q_W$ is the Toomre parameter defined in \ref{eq_toomre}. $U$ is the internal energy, $E_k$ and $E_m$ the kinetic and magnetic energies defined in Section \ref{diagnostics_av}. $H_{xy}$, $G_{xy}$, $M_{xy}$ are defined in Section \ref{diagnostics_av} and correspond to the Reynolds, gravitational and magnetic stresses. $\alpha$ is the transport coefficient defined in \ref{def_alpha} and $\dot{m}_w$ is the vertical mass loss rate {per unit area}.}  
\label{table2}
\end{table*}  
\end{center}

\subsection{Gravito-turbulent dynamo in ideal MHD}\label{ideal_mhd}

Having established consistency between the two codes and determined a converged state of hydrodynamical gravito-turbulence, we now inject an initially weak toroidal seed field into the steady state and evolve the system using ideal MHD, i.e., without ambipolar diffusion. In this case, magnetic diffusion is purely numerical and occurs on the grid scale. The geometry of the initial field is set by equation \eqref{eq_Byinit} {with $\beta_{y_0}\geq 10^4$.}
This large value of $\beta_{y_0}$ ensures that
the Lorentz force is negligible during the first few hundred $\Omega^{-1}_0$, at least on large scales, and thus any dynamo action is purely kinematic. We focus first on the $T^4$ cooling runs (AT-ZNF-ideal and PL-ZNF-ideal).

We find that the magnetic field grows rather slowly, on timescales of $250\Omega^{-1}_0$ ({\tt PLUTO}) to $500\Omega^{-1}_0$ ({\tt Athena++}) before reaching a quasi-steady state. In saturation, the magnetic energy remains smaller than the kinetic energy, with $E_m = 0.047 \langle U \rangle $ for {\tt PLUTO} and $E_m = 0.074 \langle U \rangle$ for {\tt Athena++}.  The ideal MHD run obtained with {\tt Athena++} also supports a greater Maxwell stress than the {\tt PLUTO} run. The discrepancy has two possible origins. First, the dynamo saturation and growth rates are likely sensitive to the amount of numerical dissipation (see \citetalias{riols19b}), and the time-integrator and the orbital advection algorithm are different between the two codes. In particular, the ideal MHD saturated state depends on resolution (see \S\ref{sec_res_boxsize}) and on the numerical scheme used to reconstruct the EMF at the cell edges (each scheme producing different amount of numerical dissipation, see Appendix \ref{appendixB}). Second, we do not integrate the simulations long enough for the quantities to be statistically meaningful. In particular the magnetic energy $E_m$ undergoes significant and aperiodic fluctuations on long timescales ${\sim}1000\Omega^{-1}$, suggesting that the uncertainty related to a finite averaging window could be large. However we emphasise that this bursty behaviour concerns only the ideal runs; when including ambipolar diffusion, the magnetic energy much more regular and smooth in time.

\begin{figure*}
    \centering
    \includegraphics[width=0.9\textwidth]{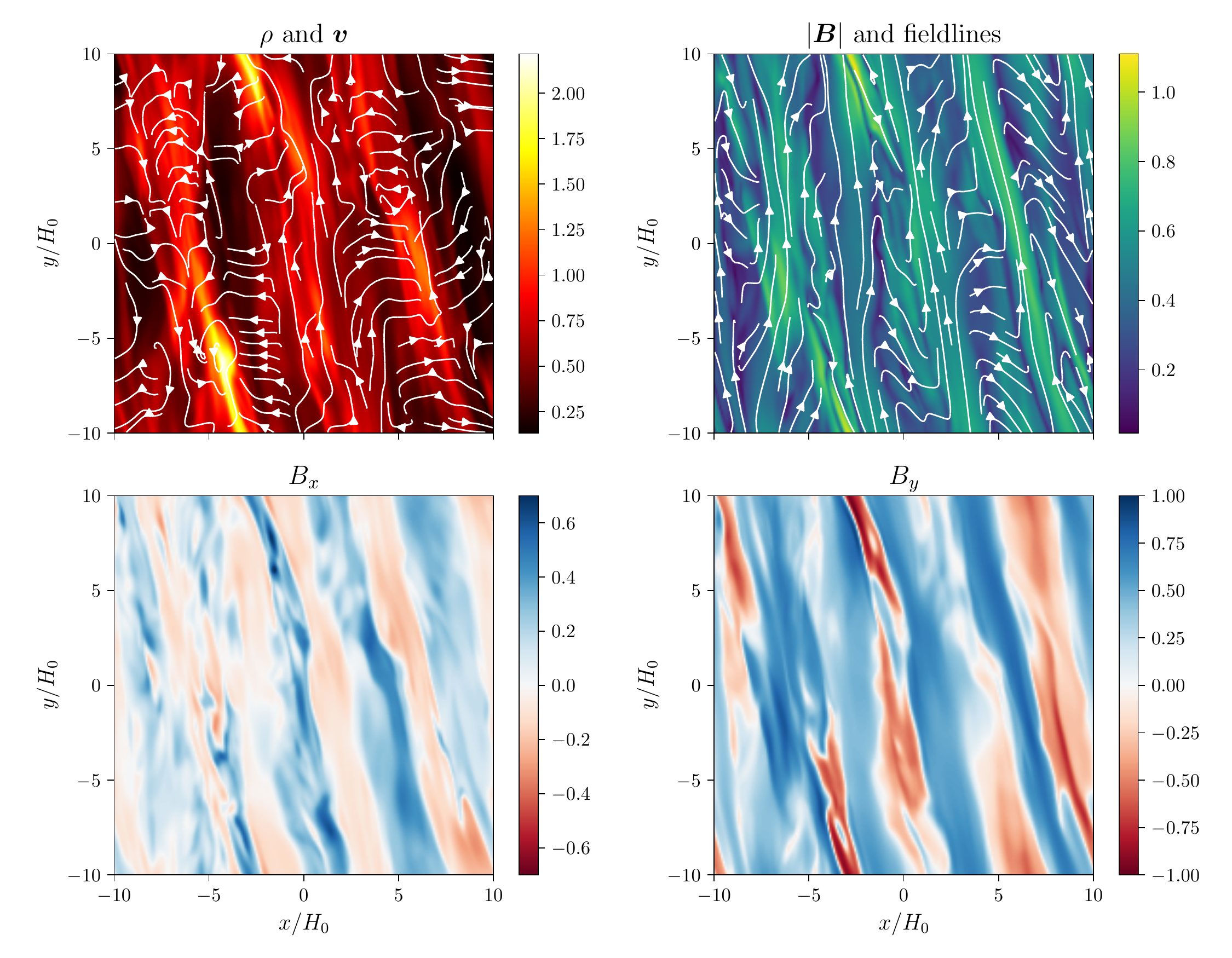}
    \caption{Snapshots taken at $z=0$ for AT-ZNF-Am10 at an arbitrary time in steady state. The dominant spiral structures have $n_x\approx 3$--$4$.}
    \label{fig_Am10_xy}
\end{figure*}

\begin{figure}
\centering
\includegraphics[width=\columnwidth]{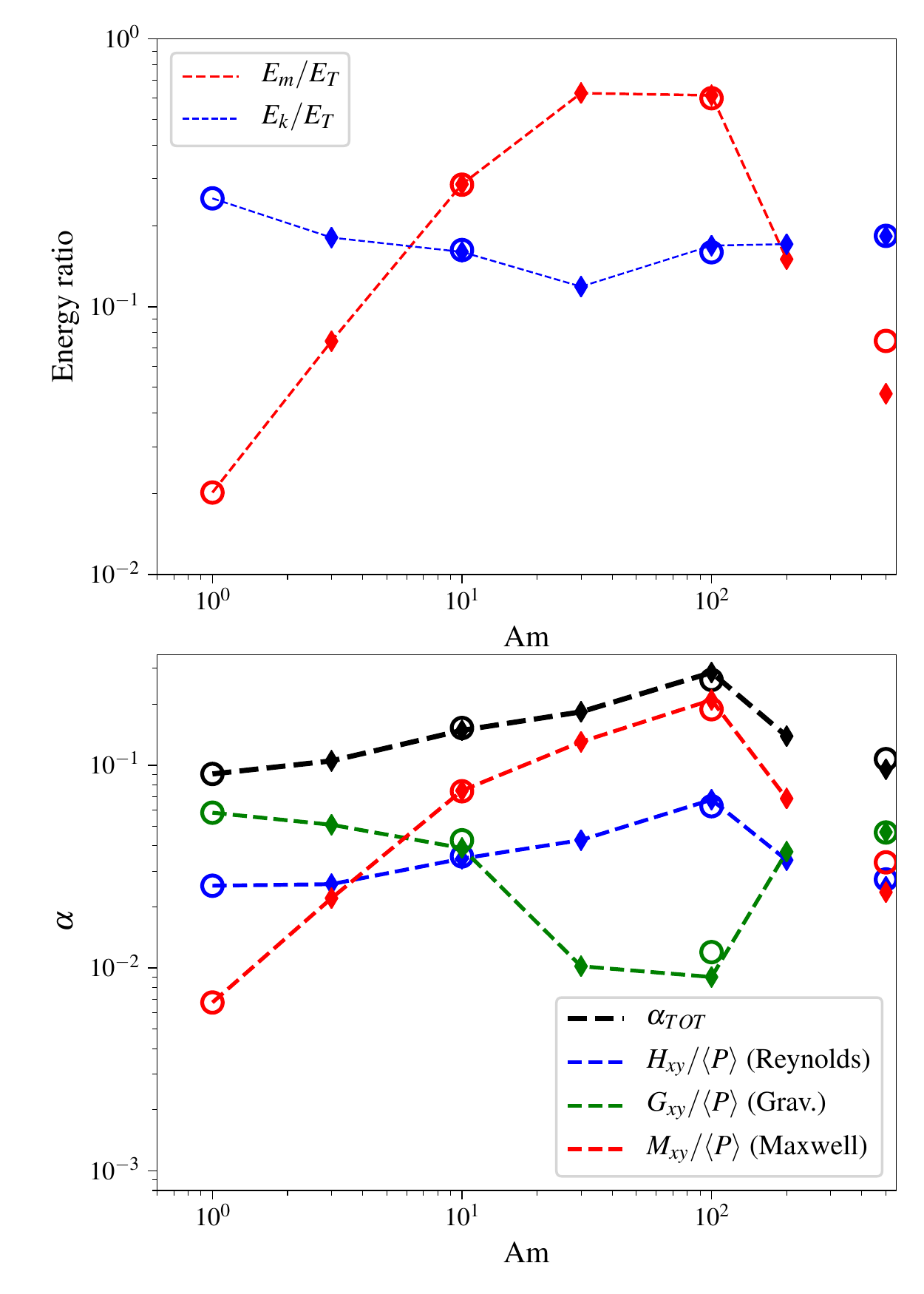}
 \caption{Top: Box averaged magnetic energy (red) and kinetic energy (blue), normalised to the averaged thermal energy, as functions of Am for $\tau_0=10\Omega^{-1}$. Bottom: Transport efficiency $\alpha$ as a function of Am. All quantities are measured in the saturated state. The diamond and circle markers correspond, respectively, to the {\tt PLUTO} and {\tt Athena++} runs. For comparison, we plot at the right edge of each panel the same quantities for the ideal case (without explicit ambipolar diffusion).}
\label{fig_saturation}
 \end{figure} 
 
 \begin{figure}
\centering
\includegraphics[width=\columnwidth]{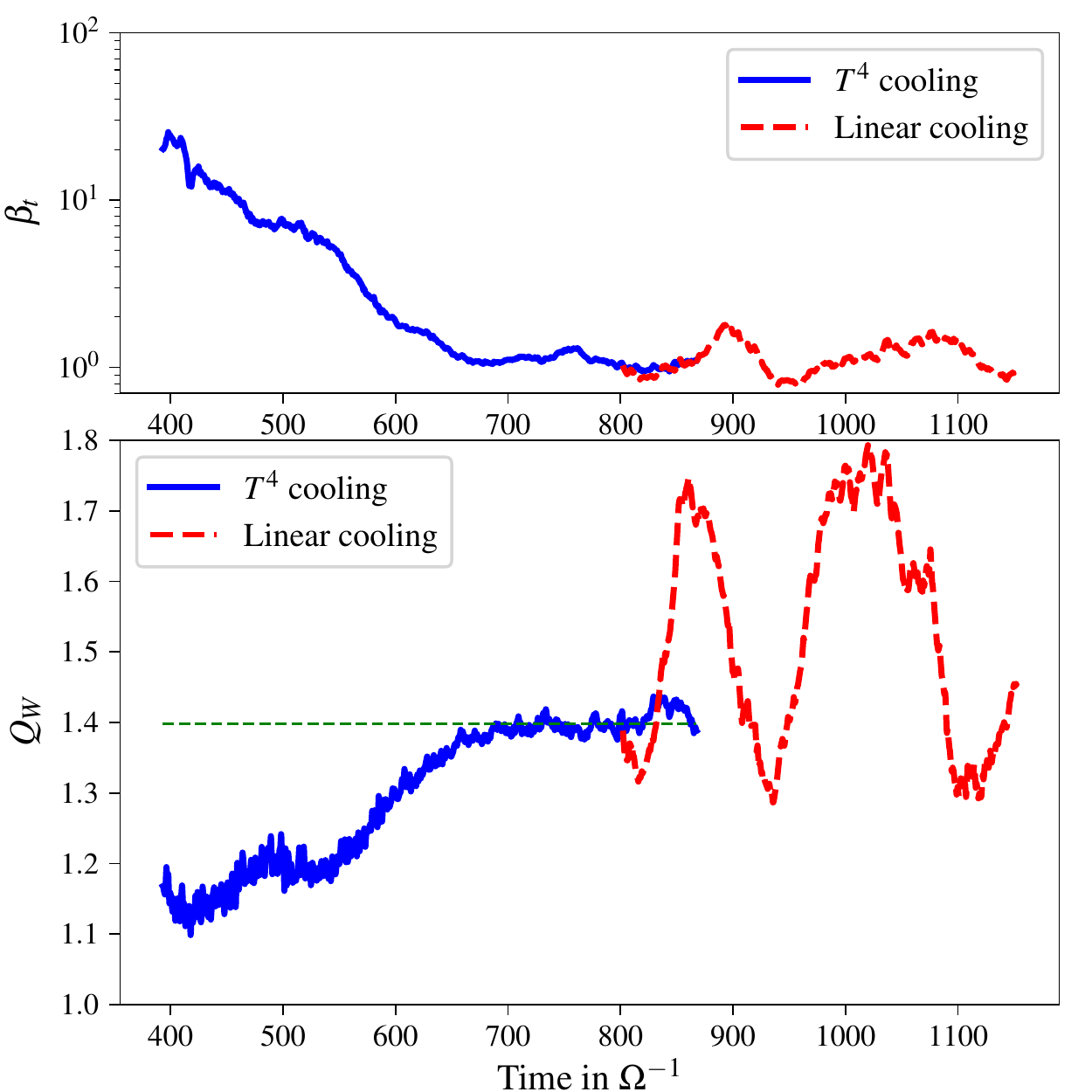}
 \caption{Time-evolution of turbulent beta plasma $\beta_t=\langle P \rangle/ E_m$ (top) and Toomre parameter (bottom) for $\text{Am}=100$. Blue/solid curves are for the $T^4$ cooling run with $\tau_0=10\Omega^{-1}$ (PL-ZNF-Am100), while the red/dashed curves are for the linear cooling run (PL-ZNF-Am100-lin), restarting from a $T^4$ cooling state.}
\label{fig_toomre}
 \end{figure}

\subsection{Gravito-turbulent dynamo with ambipolar diffusion}

We now explore the behaviour of the dynamo when ambipolar diffusion is included. Unlike Ohmic dissipation, ambipolar diffusion is a non-linear process with an associated electric field that is cubic in $\mathbf{B}$. Therefore, we expect that in the kinematic regime (at least when ${B}$ is very weak) ambipolar diffusion plays a negligible role in initial growth of the magnetic field, compared to numerical diffusion, and the results will be almost identical to those of ideal-MHD. Thus the calculation of growth rates during the kinematic phase, such as performed by \citetalias{riols19b}, is not meaningful in the presence of ambipolar diffusion only. Instead, we focus on the level of magnetic energy attained at saturation. To calculate these saturated states, we start the simulation either from the ideal MHD state or from an already existing state having ambipolar diffusion. In this way, we scan a large number of Am (ambipolar Elssaser number) ranging from 1 to 200. Table \ref{table2} summarises the different box-averaged quantities obtained at saturation for various Am. 

Before going into the box average analysis, we start by showing a view of the flow and the topology of the magnetic field.  Fig.~\ref{fig_Am10_xy} shows a snapshot of one of the simulations with $\text{Am}=10$. We see that magnetic structures reside at large scales and are correlated with the spiral density waves (we will return to this point later in Section \ref{sec_spirals}). 

We now first focus on the $T^4$ cooling runs for which $\tau_0=10\Omega^{-1}$. Figure \ref{fig_saturation} (top) shows the magnetic energy $E_m$ and kinetic energy $E_c$ (normalised to thermal energy $\langle U \rangle$) as a function of Am. Circular markers correspond to {\tt Athena++} runs, while diamond markers correspond to {\tt PLUTO} runs. When explicit diffusion is added, both codes agree well with one another, suggesting that numerical diffusion play a lesser role than in the ideal case (even for $\text{Am}=100$). As discussed later in Section \ref{sec_res_boxsize}, results are weakly dependent on resolution, although the grid is probably influencing the small-scale dynamics (especially for $\text{Am}=100$). 

We see that the dynamo is strongest when $\text{Am}=30$ and $\text{Am}=100$, for which we have $E_m \sim 0.6 \langle U \rangle $. In this regime the magnetic energy is typically $3$--$5$ times larger than the turbulent kinetic energy. This corresponds to a regime of super-equipartition fields, for which we expect important feedback of the Lorentz force on the gas flow. Indeed, the kinetic energy is reduced by a factor of almost 2 compared to the case of unmagnetized gravito-turbulence, while the gravitational stress is reduced by a factor 8 (see bottom panel of Fig.~\ref{fig_saturation}). However Fig.~\ref{fig_saturation} (bottom) shows that the transport coefficient $\alpha$ is maximal when $\text{Am}=100$, for which it is three times larger than in the hydrodynamic case. Energy is mainly extracted from the shear through the Maxwell stress in this regime. 

The drop in gravitational stress and kinetic energy is mainly because GI is weakened by strong magnetic fields. One way this is occurs is through magnetic tension; tangled magnetic fields can disrupt the hydrodynamical flow and in particular slow down the motions associated with spiral waves. Another way is by decreasing the GI growth rate: the \emph{effective} Toomre parameter, including magnetic pressure, is 
\begin{equation}
Q_{\text{eff}}\approx Q_{\text{hydro}}  \sqrt{1+\dfrac{1}{\beta_t}} ,
\end{equation}
where $\beta_t= \langle P \rangle / E_m$ \citep{kim2001}. This is $40\%$ greater than the hydrodynamic $Q$ for $\beta_t \sim 1$. Moreover, in the presence of magnetic fields, there is an increase of temperature and, consequently, an increased $Q$. To illustrate this effect, we plot in Fig.~\ref{fig_toomre} the evolution of the box-average Toomre parameter $Q_W$ as a function of time for $\text{Am}=100$. 
As the field is amplified and reaches $\beta_t \sim 1$, $Q_W$ increases by $20\%$ for $T^4$ cooling (and even by a larger factor for linear cooling, see next section). By combining the effect of magnetic pressure and the rise of thermal pressure, the effective Toomre parameter in the highly magnetized state (in particular $\text{Am}=100$) can be 75\% higher than in pure hydrodynamics, causing a drastic reduction of GI activity. 

Figure \ref{fig_saturation} indicates that the dynamo is weakened for Am $\lesssim 10$. When $\text{Am}=1$ the magnetic energy has been reduced by a factor 30 compared to $\text{Am}=100$. Because ambipolar diffusion depends non-linearly on $\bb{B}$ it is possible that the dynamo survives no matter how small Am, but will support an infinitely small amount of magnetic energy at saturation. In the other limit, for $\text{Am} \gtrsim 100$, we see from Fig.~\ref{fig_saturation} that the strength of the dynamo is reduced as Am is increased. 
{This behaviour is reminiscent of simulations with Ohmic dissipation, and is due to the kink instability developing at small scales, which can weaken the dynamo (see discussion in \citetalias{riols19b}).}

To summarise, our results show that there is little difference in terms of box-average quantities between discs regulated by ambipolar diffusion and those controlled by Ohmic dissipation \citepalias{riols19b}. The dynamo strength peaks at intermediate $\text{Am}\sim 30$--$100$. To be more quantitative and allow comparison with Ohmic runs, we calculate the effective magnetic Reynolds number defined as :
\begin{equation}
\text{Rm}_\text{eff}= \left\langle  \dfrac{\Omega H^2}{\eta_{\rm A} } \right\rangle=\left\langle \dfrac{c_{{\rm s}0}^2}{ v_{\rm A}^2 } \right\rangle \text{Am} 
\end{equation}
When $\text{Am}=30$ and $\text{Am}=100$, we find $\text{Rm}_\text{eff}\sim 0.8 \text{Am}$.  For comparison, dynamo activity was observed to be strongest when $\text{Rm}\sim 20$--$100$ in the case of Ohmic dissipation \citepalias{riols19b}. The dynamo behaves like a slow dynamo, with magnetic energy decreasing with increasing Am above a critical value. {However, we emphasise that we may not be able to probe the behaviour of the dynamo accurately at larger Am at our fiducial resolution, because for $\text{Am}\gtrsim 200$ the minimum length scale of magnetic perturbation that is not strongly damped by ambipolar diffusion (i.e., damping rate at scale $\lambda$, $\eta_A/\lambda^2$, is ${\lesssim}\Omega$) becomes comparable to or smaller than the grid scale. It is likely that, in this case, the grid (rather than explicit diffusion) controls the small-scale dynamics of the field.}


\subsection{Dependence on the cooling law}
\label{cooling_law_dep}

\begin{figure*}
    \centering
    \includegraphics[width=0.49\textwidth]{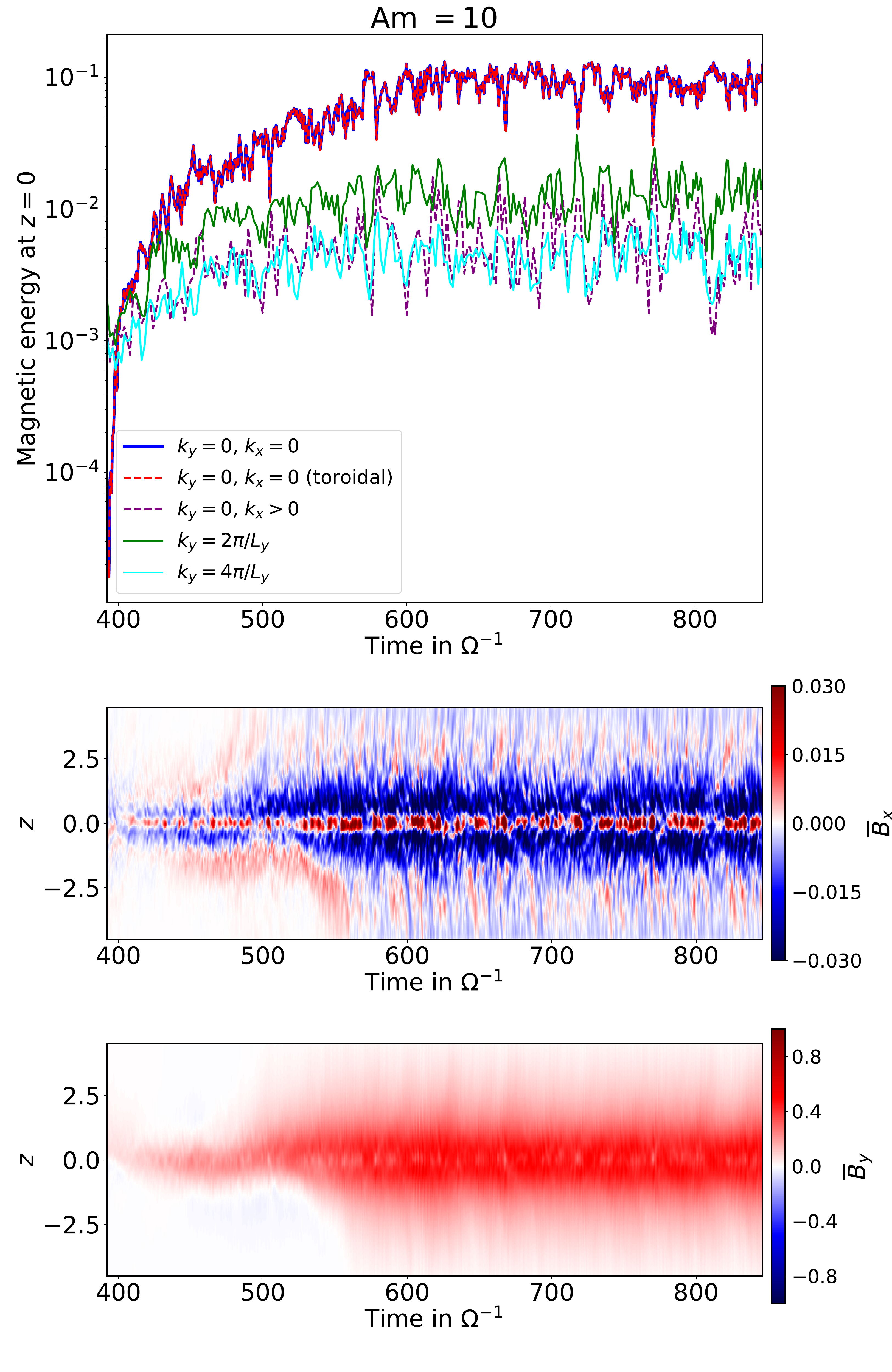}
    \includegraphics[width=0.49\textwidth]{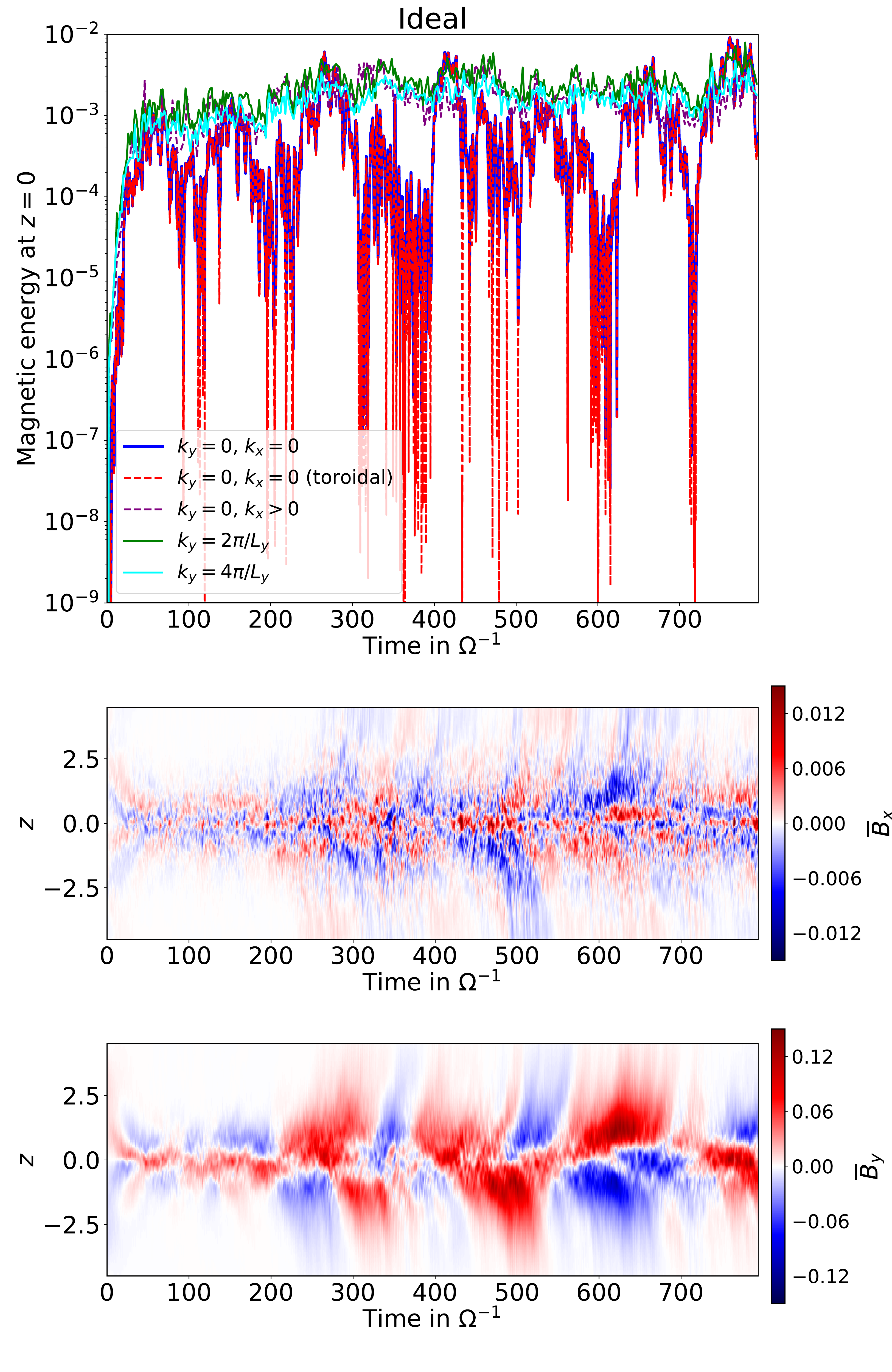}
    \caption{Top: time evolution of magnetic energy contained within the midplane ($z=0$), projected onto different Fourier harmonics, for the $T^4$ cooling law with $\tau_0=10\Omega^{-1}$. Blue/solid lines correspond to the magnetic energy of the mean field ($k_x=0, k_y=0$), while red/dashed lines corresponds to its toroidal projection. Purple/dashed lines represent the magnetic energy projected onto the axisymmetric modes ($k_x\neq0$,  $k_y=0$). Green and cyan solid lines correspond to the energy of non-axisymmetric components, respectively $k_y=2\pi/L_y$ and $k_y=4\pi/L_y$.  Lower panels: space-time $(t,z)$ diagram of the mean $\overline{B}_x$ and $\overline{B}_y$. The left panels are for $\text{Am}=10$ (run PL-ZNF-Am10) and the right panels are for the ideal case (run PL-ZNF-ideal).}
\label{fig_fftmode}
\end{figure*}

So far we have analysed the MHD simulations with $T^4$ cooling. In this section, we study how those results depend on the cooling law and make comparison with the linear cooling runs. We start with the case of ideal MHD, without explicit diffusion. The cooling time $\tau_c$ in {\tt Athena++} is chosen so that it is equal to the average effective cooling timescale $\tau_{\rm eff}$ measured in the ideal $T^4$ cooling run, rounded off to the next 0.5. We found that the magnetic energy is twice larger when a linear cooling is assumed, and it exceeds the kinetic energy (see Table \ref{table2}). As explained in Section \ref{ideal_mhd}, this difference may be due to the relatively short averaging window (note also that the linear and $T^4$ cooling runs are not averaged on the same timescale). In {\tt PLUTO} the cooling time is chosen to be equal to the effective $\tau_{\rm eff}$ corresponding to the hydrodynamical $T^4$ run, which appears to be close to the effective $\tau_{\rm eff}=10.56 \Omega^{-1}$ measured in the ideal MHD run. Table \ref{table2} shows that the magnetic energy and turbulent properties (stress, $Q_W$, etc.) at saturation depend weakly on the choice of the cooling law.

We then compare $T^4$ and linear cooling when explicit ambipolar diffusion is added. A comparison is undertaken for $\text{Am}=10$ and $\text{Am}=100$ (see  Table \ref{table2}). The cooling time $\tau_c$ is again chosen so that it is equal to the average $\tau_{\rm eff}$ measured in the corresponding $T^4$ cooling run. For $\text{Am}=10$ we then have $\tau_c=7\Omega^{-1}$ and for $\text{Am}=100$ we have $\tau_c=4\Omega^{-1}$. Table \ref{table2} shows that there is a good agreement in terms of magnetic energy and stress between the $T^4$ and the linear cooling runs. One notable difference is perhaps for $\text{Am}=100$ where $Q_W$ is slightly larger (and consequently $G_{xy}$ smaller) when a linear cooling law is used.

In addition to that, Fig.~\ref{fig_toomre} (bottom) shows that the Toomre parameter exhibits strong oscillations in the case of linear cooling, which are not present in the $T^4$ cooling run. These oscillations are correlated with the variations of the magnetisation (or $\beta_t$) and were interpreted in \citet{riols17c} and \citetalias{riols19b} as a thermal cycle involving a switch on and off of the GI activity. One possibility is that this cycle is initiated by a thermal instability, which is suppressed by the steep dependence of the $T^4$ cooling law. Further investigations will be necessary to understand the nature of this cycle.

\subsection{Fourier modes and butterfly diagrams}
\label{mean_field_components}

As in \citetalias{riols19b}, we checked whether the GI dynamo in the presence of ambipolar diffusion preferentially amplifies large or small-scale fields. For that purpose, we show in Fig.~\ref{fig_fftmode} (top left panel) the evolution of magnetic energy associated with different Fourier modes in the runs PL-ZNF-Am10 ($\text{Am}=10$, $T^4$ cooling).  The calculation is restricted to the midplane $z=0$, as it contains most of the magnetic energy. Clearly the dominant component of the magnetic field is the $k_x=k_y=0$ mode (mean field $\mathbf{\overline{B}}$), which is essentially toroidal. The dynamo in this regime is thus large scale. We see that the next important component is the $k_y=2\pi/L_y$ mode associated with the large scale spiral waves. Although significantly smaller than the mean field (its energy being about 8 times smaller), this component is necessary for the large scale dynamo process. As we will show, it contributes to the turbulent EMF and helps to regenerate the mean field against ambipolar diffusion. Finally Fig.~\ref{fig_fftmode}  suggests that the axisymmetric modes ($k_y=0$) with $k_x>0$ are quite negligible in the magnetic budget. 

In the lower left panels of Fig.~\ref{fig_fftmode}, we show the time-evolution of the vertical profiles of the mean magnetic field $\overline{B}_x(z)$ (centre) and $\overline{B}_y(z)$ (bottom). The large-scale field remains confined within the midplane region of the disc and is mainly toroidal with $\overline{B}_y/\overline{B}_x \simeq 20$ at $z = \pm H_0$. Both poloidal and toroidal components have the same polarity in the midplane, but $\overline{B}_x$ changes its sign at $z \simeq \pm  0.2 H_0$. Note that the evolution of the radial field show strong variations in the midplane which are probably associated with the intermittent nature of GI activity.  Finally an important feature is that the large-scale field does not flip (the polarity remains fixed) and is symmetric about the midplane (i.e., it has an `even' geometry). All these properties are similar to those obtained in simulations with explicit Ohmic resistivity \citepalias{riols19b}. Note that we obtain a similar field geometry for all zero-net flux simulations with ambipolar diffusion (even for $\text{Am}=200$). Also the profile of $\overline{B}_x(z)$ and $\overline{B}_y(z)$ remains similar when increasing the vertical box size (see Section \ref{sec_res_boxsize}). \\

As a comparison we computed the same diagnostics for the ideal run PL-ZNF-ideal. Fig.~\ref{fig_fftmode} (bottom right) shows that the mean field flips many times during the course of the simulation but its evolution is irregular (it can be either anti-symmetric or symmetric about the midplane). The period of the flipping is about $200 \Omega^{-1}$ in the saturated state. Fig.~\ref{fig_fftmode} (top right) shows that, unlike the case $\text{Am}=10$, the mean component of the field $\mathbf{\overline{B}}$ in the midplane is not the dominant one.  Even when it is symmetric about the midplane (or at least when it is not changing polarity at $z=0$), its energy is comparable to the non-axisymetric mode $k_y=2\pi/L_y$. 

\begin{figure*}
    \centering
    \begin{flushright}
    \includegraphics[width=.6\textwidth]{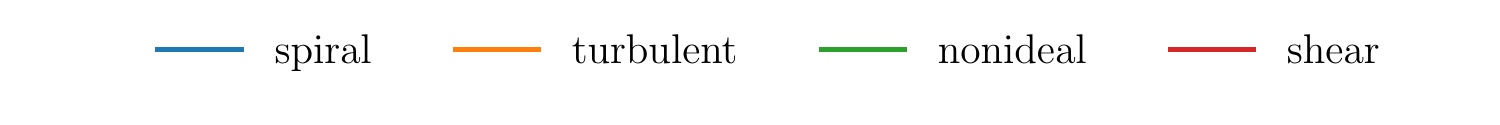}
    \end{flushright}
    \vspace{-1.5em}
    \includegraphics[width=\textwidth]{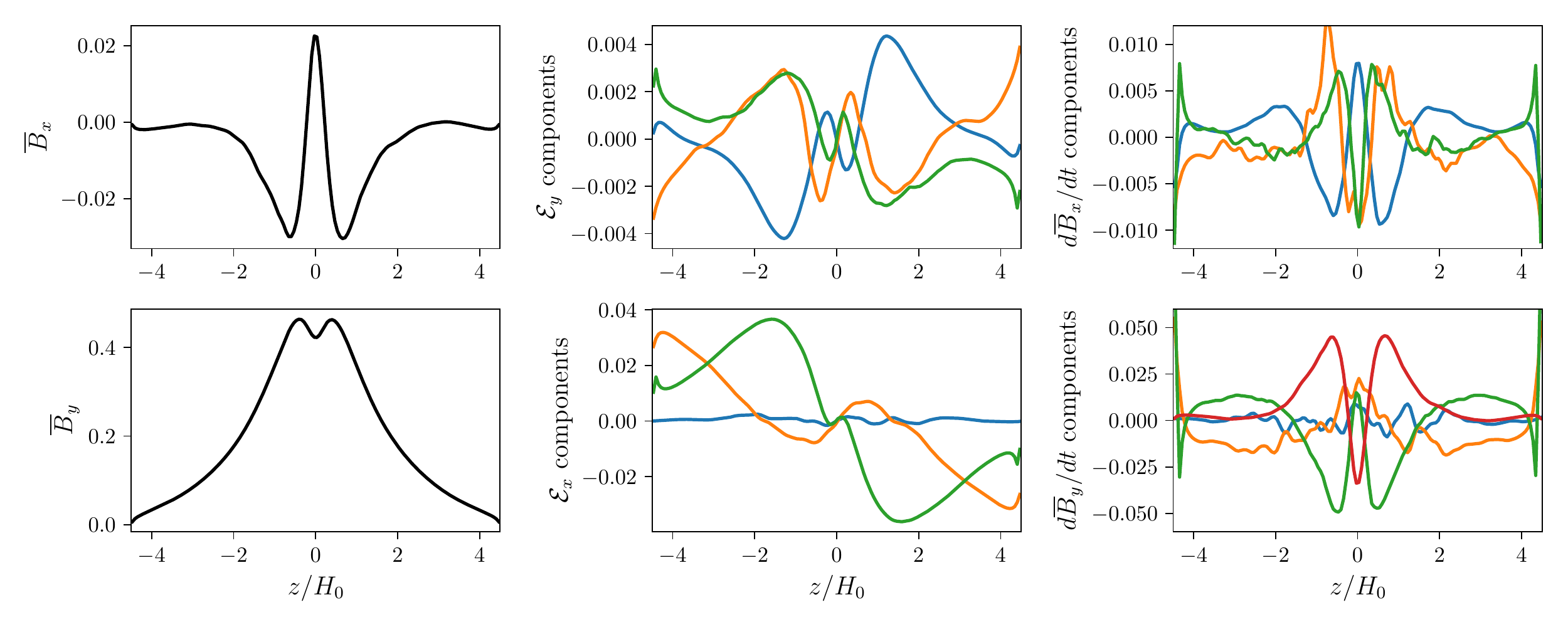}
    \caption{Time-averaged vertical profile of mean magnetic field $\bar{B}_x,\bar{B}_y$; mean electrmotive force $\mathcal{\overline{E}}_x,\mathcal{\overline{E}}_y$; and $\rmd\bar{B}_x/\rmd t$, $\rmd\bar{B}_y/\rmd t$ for run AT-ZNF-Am10 in its steady state. The behaviours are characteristic of runs with relatively high Am. We separate the electromotive force into several components: $\mathbf{v}\times\mathbf{B}$ in the $n_x=4, n_y=1$ phase folded profile, which mainly accounts for the contribution from $n_x=4, n_y=1$ spirals waves (blue curves); the remaining $\mathbf{v}\times\mathbf{B}$, which mainly accounts for turbulent electromotive force (orange curves); and the non-ideal electromotive force, which in this case accounts for ambipolar diffusion. We use the same colour code for $\rmd\bar{\mathbf{B}}/\rmd t$ due to these components of the electric field, and the red curve in the $\rmd B_y/\rmd t$ panel corresponds to the generation of $\bar{B}_y$ through the shearing of $\bar{B}_x$. The divergences seen at the boundaries are probably due to the boundary conditions, but they do not affect the dynamo process, which happens mainly near the midplane (see Section \ref{subsec_box_size}).
 }
    \label{fig_EMF}
\end{figure*}

\section{Generation of mean field by spirals}
\label{sec_spirals}

In this section we investigate how the mean field is generated and sustained by GI motions. Our focus will be on the role of spiral density waves; as has been shown by \citetalias{riols19b}, spirals are the key to explaining the generation of $\mathbf{\overline{B}}$ in discs regulated by Ohmic dissipation. One would like to know whether or not these structures are also important when ambipolar diffusion dominates

\subsection{EMF budget}

We show in Fig.~\ref{fig_EMF} the vertical profiles of the electromotive forces and the terms involved in the evolution of ${\overline{B}_x}$ (top) and ${\overline{B}_y}$ (bottom) for $\text{Am}=10$ and $T^4$ cooling law (run AT-ZNF-Am10). This includes the spiral and turbulent part of the mean ideal electromotive force   ($\overline{\bb{\mc{E}}}^\star$ and $\overline{\bb{\mc{E}}}^t$ defined in Section \ref{sec_MF}), as well as the shear stretching (the `omega effect') and the ambipolar EMF   $\overline{\mathbfcal{E}}^{NI}(z)$ (see equations \ref{eq_Bxmean} and \ref{eq_Bymean}). Recall that the spiral term $\overline{\bb{\mc{E}}}^\star$ is computed using the phase-folded technique (see Section \ref{phase_folding}), using a spiral pattern with $k_{x_s}=4 k_{x_0}$ with $k_{x_0}=2 \pi / L_x$ (this corresponds to the most prominent spiral mode seen in the {\tt Athena++} and {\tt PLUTO} runs). All the terms are averaged in time, during the saturated phase ($\sim 250 \Omega^{-1}$), and over the horizontal plane. 

Let us begin with the toroidal component, ${\overline{B}_y}$. Above the midplane for $\vert z\vert >0.25 H_0$, it is clear from the bottom right panel of Fig.~\ref{fig_EMF} that ${\overline{B}_y}$ is generated mainly via the background shear flow (red line). In the central panels, the radial turbulent EMF (orange) has a negative gradient in $z$, like ambipolar diffusion (green), and therefore acts as a turbulent diffusion on the mean field. In the midplane ($\vert z \vert <0.25 H_0$), however, the configuration is different since ${\overline{B}_y}$ and ${\overline{B}_x}$ have the same polarity, and therefore the omega effect tends to destroy the large-scale field (this explains the little dip in the ${\overline{B}_y}$ profile at $z=0$). We note that the EMF budget for ${\overline{B}_y}$ is actually similar to that obtained with Ohmic resitivity (see Fig.~8 of \citetalias{riols19b}). Despite its negative contribution in the midplane, the omega effect is essential in the vertically averaged budget, and therefore a poloidal field is required for the large-scale dynamo to work. 

We then study how the mean ${\overline{B}_x}$ is produced. The top panels of Fig.~\ref{fig_EMF} shows the spiral, turbulent, and ambipolar EMFs profiles associated with the radial field $\overline{B}_x$. For $\vert z \vert <0.25 H_0$ (midplane region), the top right panel shows that $\overline{B}_x>0$ is generated mainly via the spiral EMF in $y$ (blue) and dissipated through the non-ideal ambipolar EMF (green) and the turbulent EMF (orange).  Above the midplane, for $\vert z \vert >0.25 H_0$ and $\vert z \vert \lesssim 1.5 H_0$, $\overline{B}_x<0$ and the gradients of $\overline{\mathcal{E}}^{t}_y$, $\overline{\mathcal{E}}^{s}_y$ and $\overline{\mathcal{E}}^{NI}_y$ are just reversed, meaning again that the spiral EMF helps to sustain the mean radial field against ambipolar and turbulent diffusion. \\

In summary, we have shown that the mean field dynamo relies on 
the production of $B_y$ from $B_x$ by the omega effect (primarily)
and on the generation of $B_x$ by relatively large-scale fluctuations issuing from the spiral density waves.

\subsection{Dynamo process and phase-folded profiles}

As demonstrated in the previous subsection, spiral motions are key to in regenerating the mean field $\overline{B}_x$. But how exactly do they operate? In the presence of Ohmic dissipation, \citetalias{riols19b} proposed a physical picture that highlights the role of large-scale vertical rolls associated with spiral density waves in the generation of $\overline{B}_x$ (see their sections 5.1 and 5.2). These rolls are a fundamental feature of linear density waves in stratified discs and arise from a baroclinic effect associated with the thermal stratification of the disc \citep{riols18}. The mechanism of radial field generation from a toroidal field takes place through the following steps  (see in particular Fig.~12 of \citetalias{riols19b}): 
\begin{enumerate}
    \item First, spirals horizontally compress and tilt the initially toroidal field. This creates a negative (positive) $B_x$ inside (outside) of the spiral wave. On average, no mean radial field ($\overline{B}_x$) is produced via this process.
    \item At the centre of the spiral, the field is pinched and lifted by the vertical velocity associated with the roll structure. This carries some magnetic flux to higher altitude and creates a $\tilde{B}_z$ component with opposite polarity on both sides of the spiral. 
    \item  As the rolls `turn over' above the disk at $z\sim H_0$, the field is then stretched in the radial direction and folded, producing a net $\overline{B}_x$ with opposite sign to $B_y$ in the corona (through the EMF term $\tilde{u}_x \tilde{b}_z$).
\end{enumerate}

\begin{figure*}
    \centering
    \includegraphics[width=.9\textwidth]{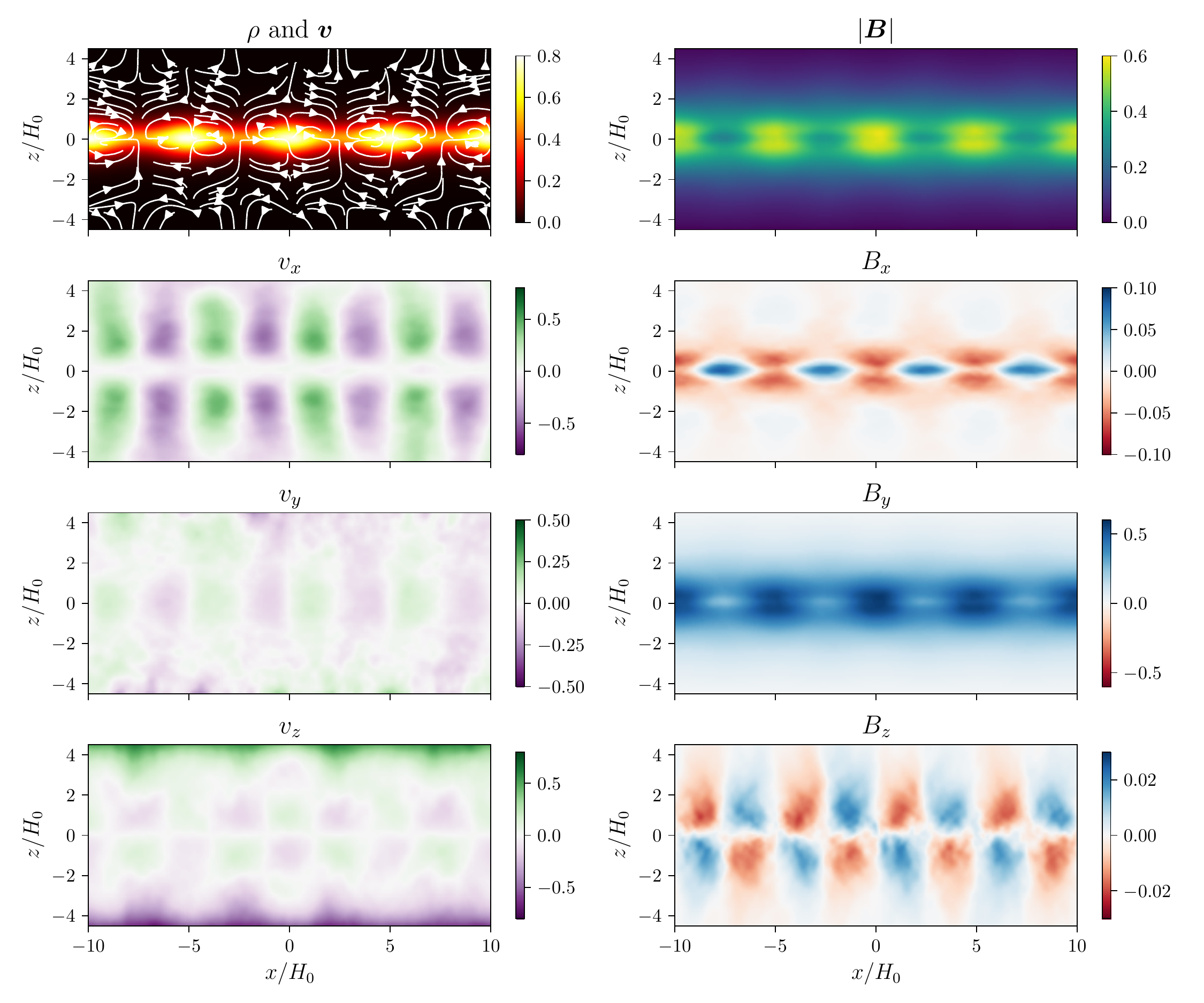}
    \caption{Phase-folded profile in the steady state of run AT-ZNF-Am10. Turbulent fluctuations are suppressed by phase-folding multiple snapshots, and the velocity rolls associated with the spiral density waves and the twisting of field lines due to the velocity rolls are clearly visible, despite their amplitudes being often much smaller than those of the turbulent fluctuations.}
    \label{fig_Am10_xz_phase_fold}
\end{figure*}

One may wonder if the mechanism described above and detailed in \citetalias{riols19b} still hold in discs dominated by ambipolar diffusion. 
To this end, we examined the gas motions and the magnetic topology around spiral waves for $\text{Am}=10$. Instead of looking at fixed snapshots, as in \citetalias{riols19b}, we compute the `mean' motion and magnetic field associated with the spirals by phase folding the data (see Section \ref{phase_folding} for details about this technique). This procedure offers a visualisation of the flow in a poloidal plane ($x$,$z$), in which perturbations associated with the most prominent spiral mode are preserved but turbulent and uncorrelated fluctuations are suppressed. The result is shown in Fig.~\ref{fig_Am10_xz_phase_fold}. The most prominent spiral mode in the {\tt Athena++} and {\tt PLUTO} runs is the mode $k_y=2\pi/L_y$ and $k_x=4k_{x_0}=8 \pi/L_x$ (see Fig.~\ref{fig_hyd}, which shows four spiral arms in the box). 
First, we clearly identify in the top {left} panel of Fig.~\ref{fig_Am10_xz_phase_fold} the pattern of four counter-rotating, large-scale rolls associated with each density maximum. These rolls are located within $z \pm H_0$, but the radial motions directed outward from the density wave persist at larger $z$ in the corona (see also $v_x$ panel). 

Figure \ref{fig_Am10_xz_phase_fold} shows that the toroidal field $B_y$ in the midplane is compressed within the spiral density maxima (see $b_y$ panel) and twisted along the direction of the spirals, which produces a negative $B_x$ component inside the spirals and a positive component outside (see $b_x$ panel). Figure \ref{fig_Am10_xy} (top right panel) shows a snapshot of the magnetic-field lines projected in the midplane and confirms this picture. This affirms step (i). The vertical motions associated with the velocity rolls of the spirals then bring the negative $B_x$ upwards into the corona (see $b_x$ panel). This leads to a mean $\overline{B}_x>0$ in the midplane and a mean $\overline{B}_x<0$ above, consistent with Fig.~\ref{fig_fftmode}. Another way to check this result is to see that, at the location of the density maxima where $B_x<0$, the vertical velocity of the rolls is always directed outwards (positive $\tilde{u}_z$ if $z>0$ and negative
if $z<0$).  In the inter-arm regions, we observe the opposite configuration. A clear correlation is then obtained between $u_z$ and $B_x$, which generates a positive gradient of $-\mathcal{E}_y$ and then a positive $\overline{B}_x$ near the mid-plane. 
The $B_z$ panel of Fig.~\ref{fig_Am10_xz_phase_fold} shows that, for $z>0$, a positive vertical field is produced on the left side of the spiral wave front and a negative $B_z$ on the other side of the wave front. This configuration can be explained simply by the fact that the field line, which crosses the spiral wave at some angle, is pinched and lifted up at the centre of the spiral by the rolls. All of these elements confirm that step (ii) is at work in our simulations. 
Finally, the $B_x$ panel of Fig.~\ref{fig_Am10_xz_phase_fold} shows 
red `mushroom-like' structure in the contours of $B_x$, reflected on either side of the midplane. These structures are similar to those observed in the snapshots of Fig.~15 of \citetalias{riols19b} and indicate that the field that reaches $z\simeq H_0$ is
stretched horizontally by the counter-rotating motions of each roll. 
We see that, in the corona, the vertical component  of the field is correlated with the rolls' horizontal motions (in regions where $u_x<0$, $B_z$ is positive, and vice versa). Such correlations lead to a negative EMF $-\mathcal{E}_y=\tilde{u}_x\,\tilde{b}_z$, resulting in a negative gradient for the EMF and therefore a negative $\overline{B}_x$, as expected from Fig.~\ref{fig_fftmode}. The physical processes described in step (iii) seem to be at work in our simulations. 

While generating $\bar B_x$ with an opposite sign to $\bar B_y$ in the corona, the dynamo mechanism above tends to generate $\bar B_x$ with the same sign as $\bar B_y$ at the midplane, which is visible in Figure \ref{fig_EMF}. This effect, together with the shear of $\bar B_x$ to generate $\bar B_y$, leads to cyclic flipping of the polarity in the absence of diffusion, and may explain the dependence of field geometry (flipping or fixed) on the level of diffusion seen in Figure \ref{fig_fftmode} and \citetalias{riols19b}.


\section{Dependence on resolution and box size}
\label{sec_res_boxsize}
\subsection{Resolution}

\begin{figure}
    \centering
    \includegraphics[width=.47\textwidth]{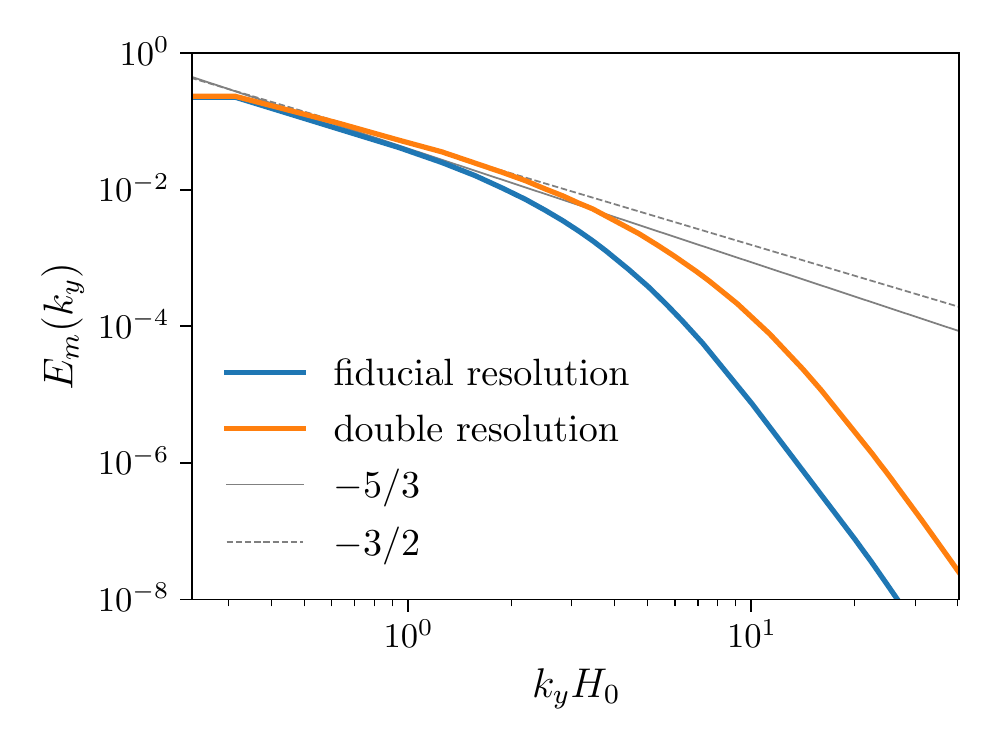}
    \caption{Spectra of magnetic energy for runs AT-ZNF-Am10 (fiducial resolution) and AT-ZNF-Am10-HR (double resolution). Increasing resolution leads to more small-scale (high ${k_y}$) fluctuations. However, the amplitudes of large-scale (low ${k_y}$) modes, which dominate the magnetic energy, are similar.}
    \label{fig_spectra_res}
\end{figure}

\begin{figure}
    \centering
    \vspace{-1.5em}
    \includegraphics[width=.5\textwidth]{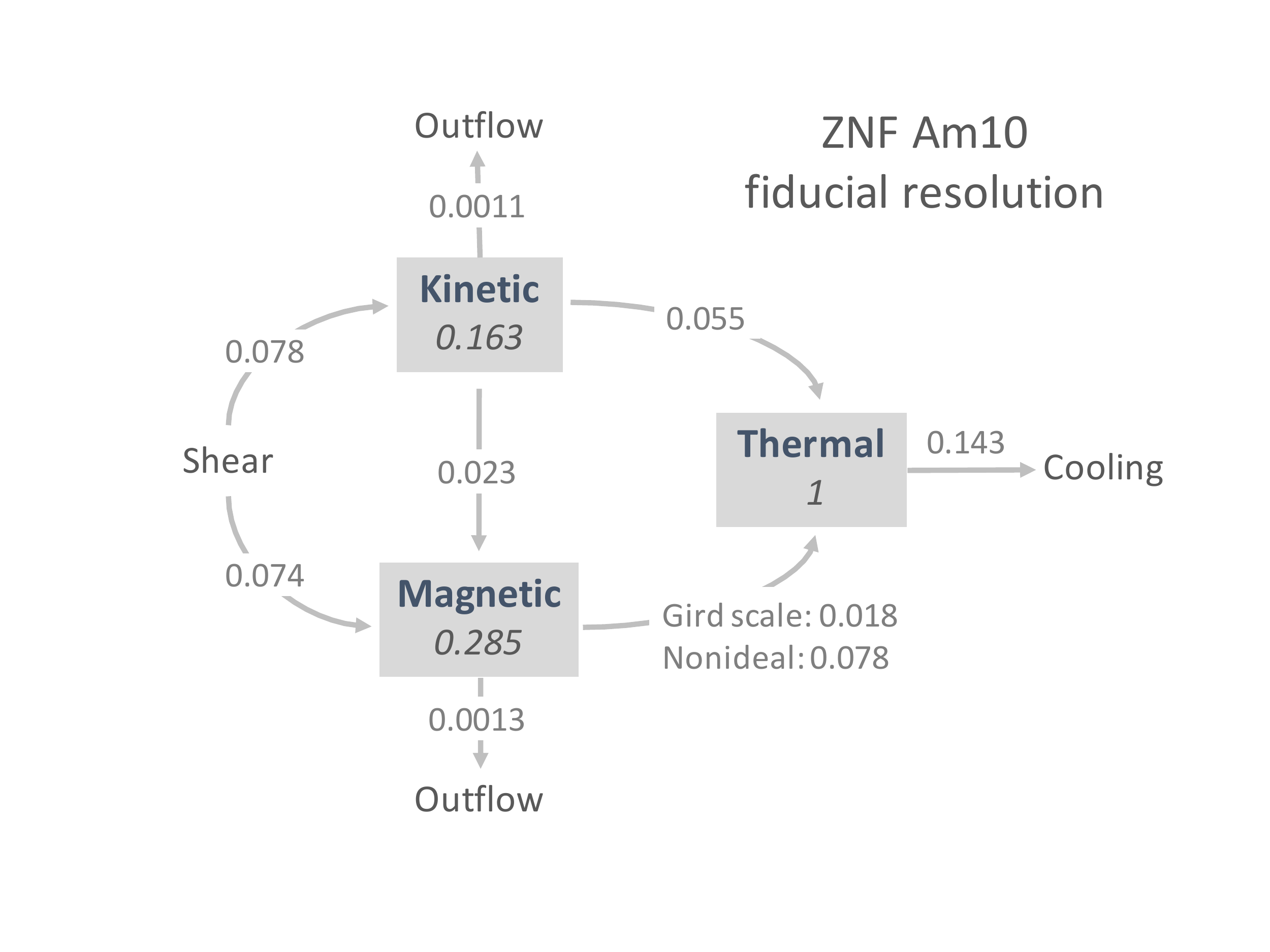}\\
    \vspace{-3em}
    \includegraphics[width=.5\textwidth]{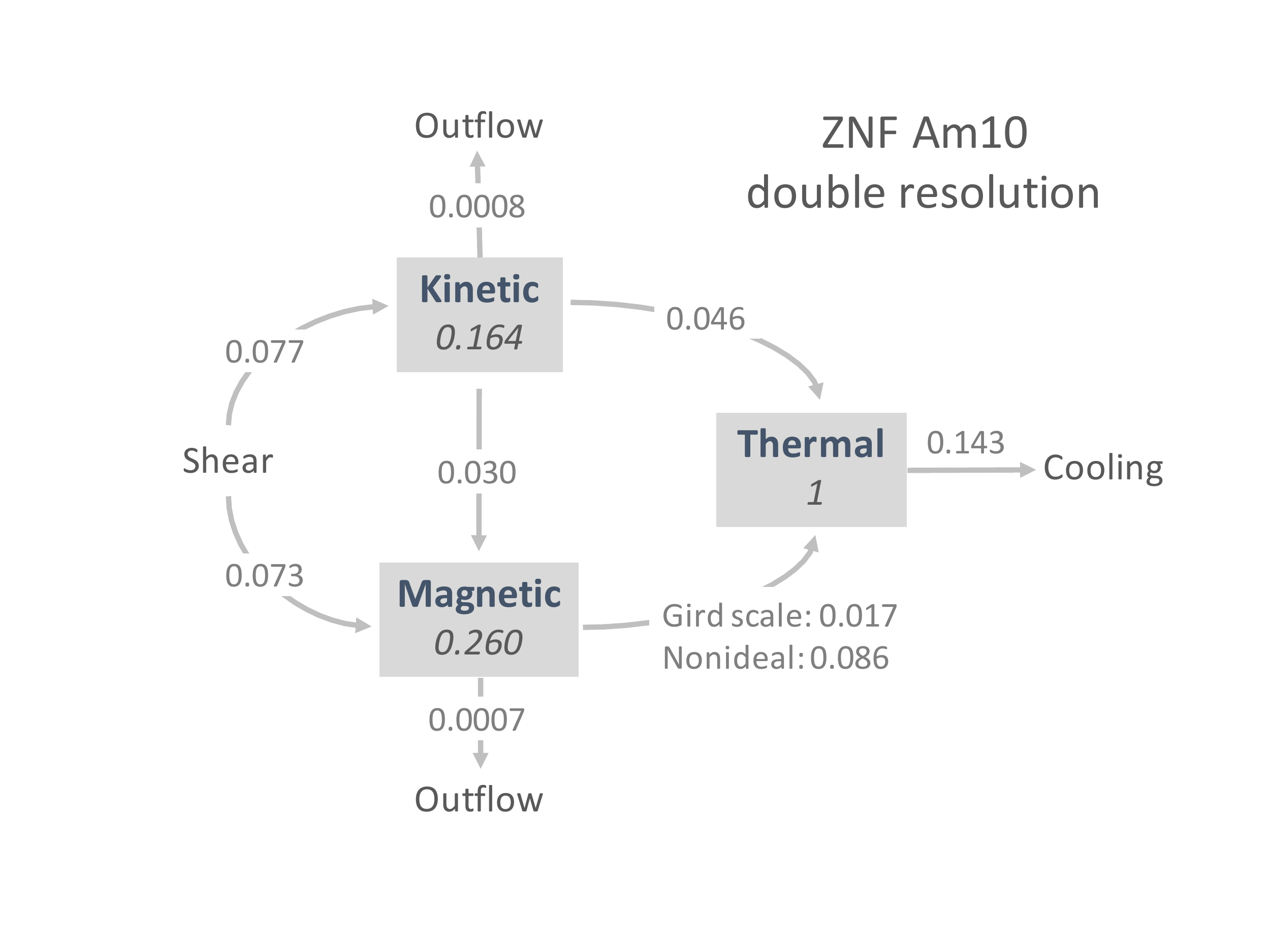}\\
    \vspace{-1.5em}
    \caption{Average rate of energy generation and transport in steady state in runs AT-ZNF-Am10 (top) and AT-ZNF-Am10-HR (bottom). The energies are normalised by the thermal energy, and energy generation and transport rates are normalised by the thermal energy $\times \Omega$. Shear generates kinetic energy through Reynolds and gravitational stresses, and magnetic energy through Maxwell stress. We split the {conversion} of magnetic to thermal energy into two channels: numerical dissipation at the grid scale and ambipolar heating.}
    \label{fig_energy_res}
\end{figure}

\begin{figure}
    \centering
    \includegraphics[width=.5\textwidth]{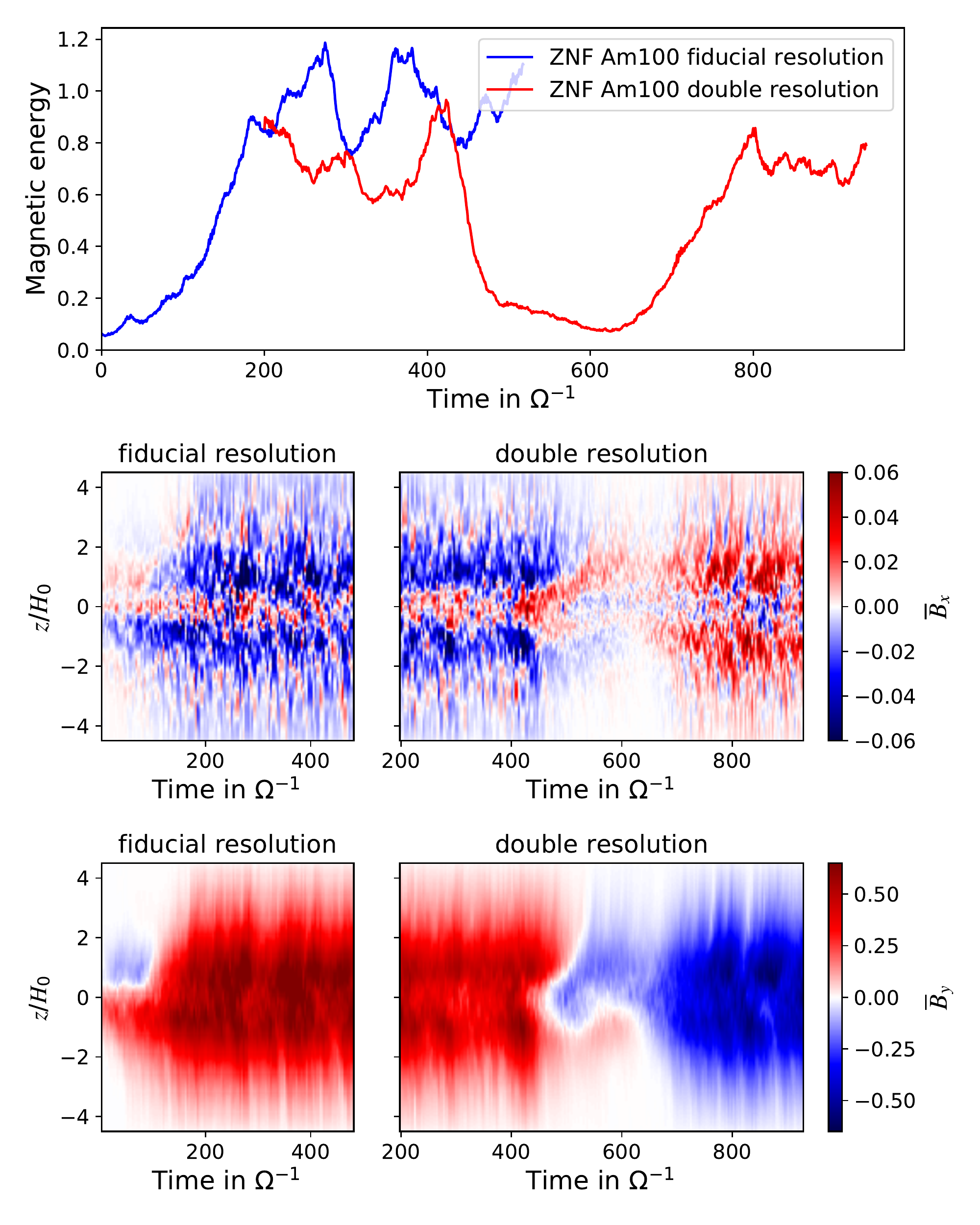}
    \caption{Evolution of magnetic energy and butterfly diagrams for $\bar{B}_x,\bar{B}_y$ in runs AT-ANF-Am100 (fiducial resolution) and AT-ANF-Am100-HR (double resolution). The large difference in magnetic energy in Table 2 is mainly because the double-resolution run contains a flip of magnetic polarity, during which the magnetic energy is low due to the lack of a strong mean field. Before and after the flip, the magnetic energy remains similar to that obtained at fiducial resolution.}
    \label{fig_Am100_res}
\end{figure}

To study whether our fiducial resolution is sufficient to capture all of the relevant dynamics, we re-perform the set of zero-net-flux, $T^4$ runs at double resolution ($512 \times 512 \times 288$) for ideal MHD, $\text{Am}=100$ and $\text{Am}=10$ with {\tt Athena++}, and $\text{Am}=100$ with {\tt PLUTO}. In {\tt Athena++}, we restart the simulations from an MHD quasi-steady turbulent state at fiducial resolution and interpolate the magnetic field so that $\grad\bcdot\bb{B}=0$ is preserved. {(Because the code stores magnetic fluxes through cell interfaces, a simple linear interpolation of magnetic flux density onto new interfaces guarantees that the total magnetic flux through a cell remains zero when doubling resolution.)} In {\tt PLUTO}, we start from a gravito-turbulent hydrodynamical state and add a strong toroidal magnetic perturbation in the midplane of the form $B_y=0.2 \exp{(-z^2/2H_0^2)}$.

When $\text{Am}=10$, the fiducial-resolution run AT-ZNF-Am10 and the double-resolution run AT-ZNF-Am10-HR show very similar box-averaged quantities (Table \ref{table2}). The magnetic energy spectra for the two runs are also very similar, although the double-resolution runs show more small-scale fluctuations (Figure \ref{fig_spectra_res}). To confirm that these additional small-scale fluctuations do not affect large-scale dynamics, we calculate the rate of energy generation and transport in the steady state for the two simulations (Fig.~\ref{fig_energy_res}). Most of the transport rates are barely affected by the increase in resolution. In particular, the similarity in the grid-scale dissipation of magnetic energy between the two resolutions, despite the different level of small-scale fluctuations, suggests that the energy cascade rate is set by large-scale processes that are relatively insensitive to the grid resolution.

When $\text{Am}=100$, the box-averaged quantities in the fiducial- and double-resolution runs (with {\tt Athena++}) appear to be quite different but, as Figure \ref{fig_Am100_res} shows, this is mainly because the double-resolution run contains a flip in the sign of the large-scale magnetic field while the fiducial run does not. If we exclude the flip from the averaging, the averaged quantities of the two runs becomes similar (the magnetic energy in the double-resolution run is ${\approx}70\%$ of that obtained at the fiducial resolution). The increased tendency for flipping, on the other hand, is probably because of the decreased numerical dissipation. Previously we showed that non-ideal diffusivity tends to suppress flipping, and it is likely that numerical dissipation can produce a similar effect.

For ideal MHD, the fiducial- and double-resolution runs give significantly different box-averaged quantities, especially for the magnetic energy. The lack of numerical convergence is to be expected, since we believe small-scale kink instabilities, which are sensitive to the level of numerical dissipation in the absence of non-ideal effects, affect the strength of the dynamo \citepalias{riols19b}.


\subsection{Vertical box size}
\label{subsec_box_size}
\begin{figure*}
    \centering
    \includegraphics[width=0.49\textwidth]{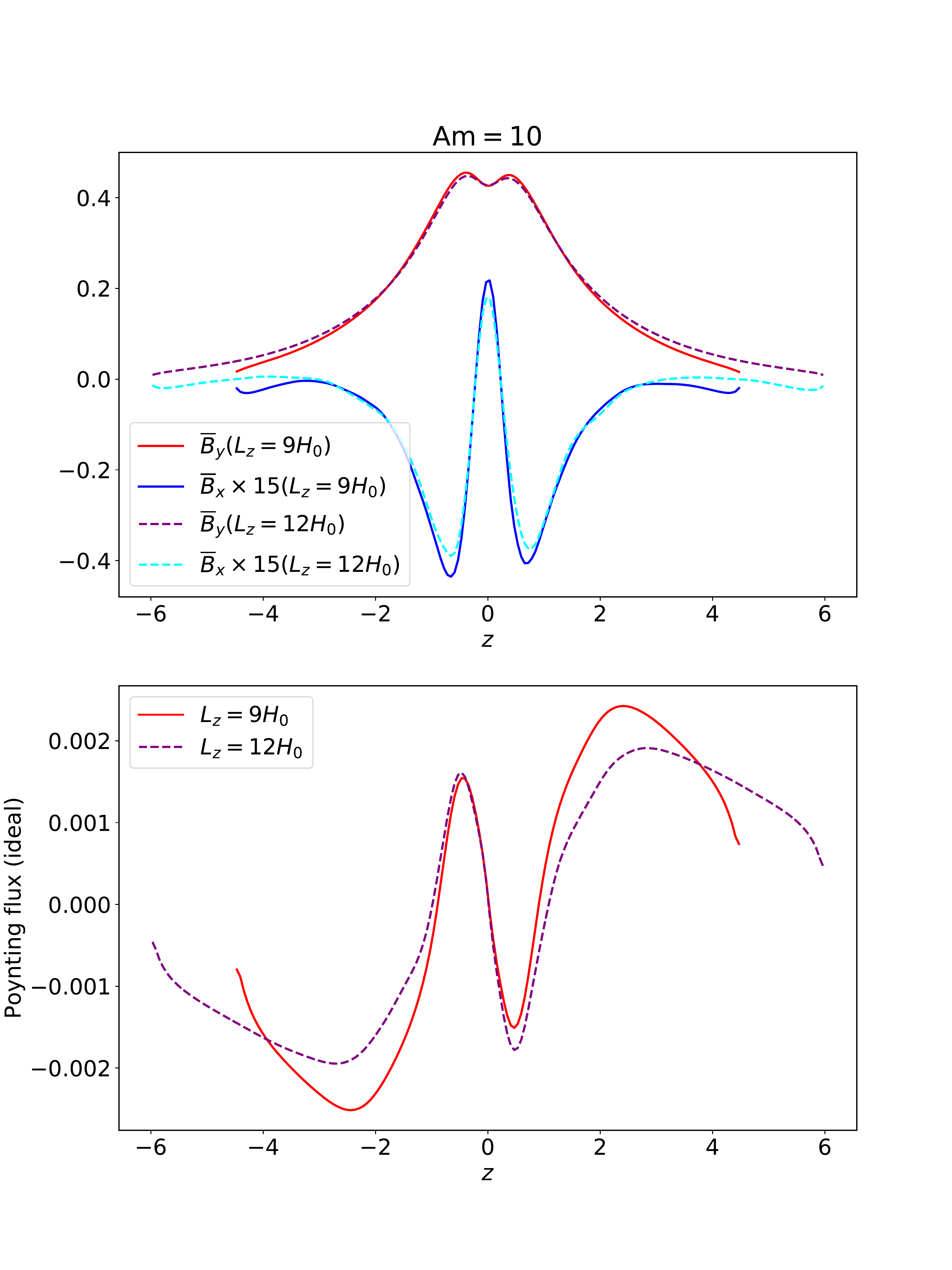}
    \includegraphics[width=0.49\textwidth]{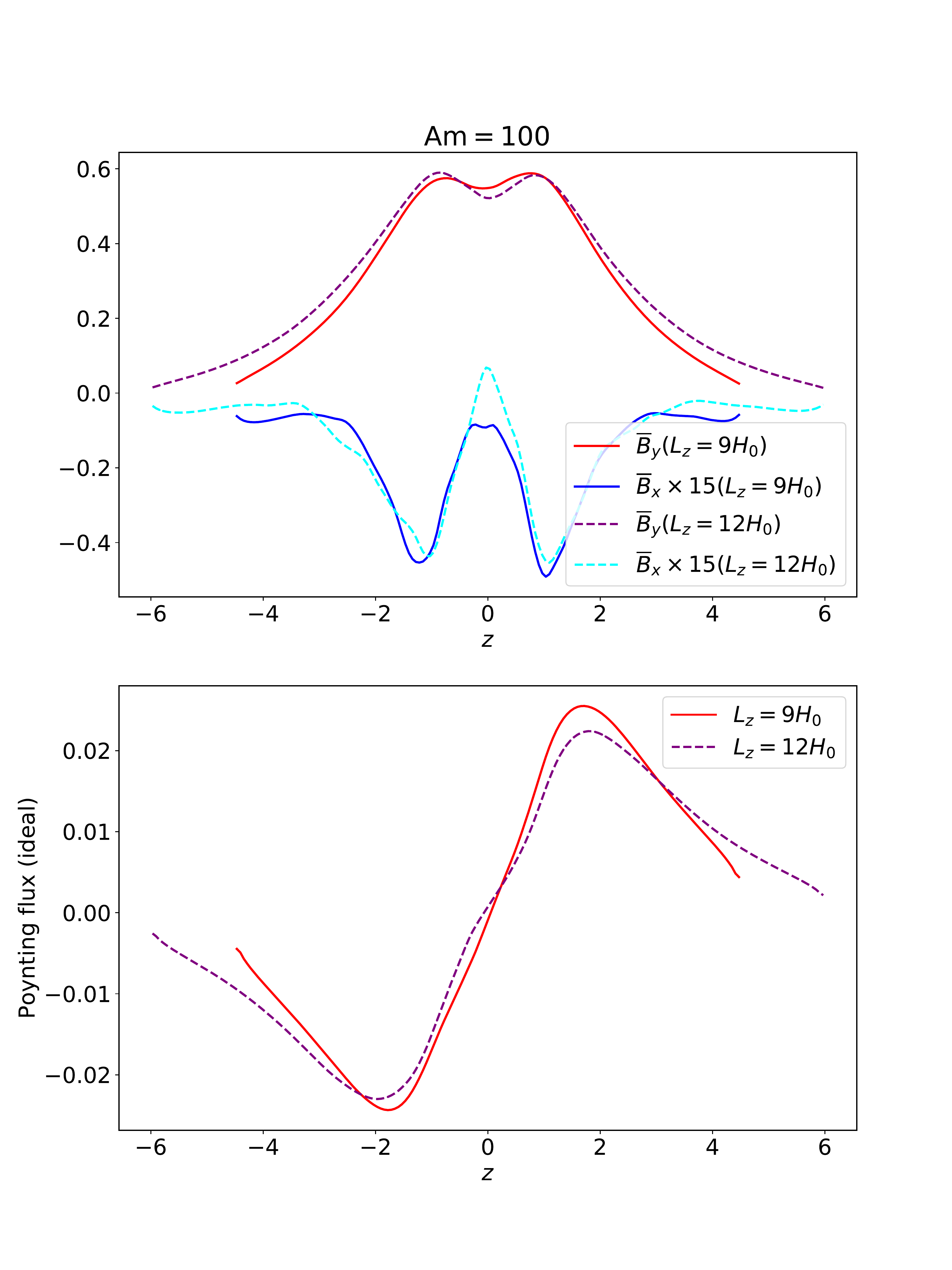}
    \caption{Top panels: vertical profiles of the mean $\overline{B}_x$ and $\overline{B}_y$ averaged in time.   Bottom panels: vertical profiles of the ideal Poynting flux $\overline{B^2 v_z - B_z (B \cdot v)}$.  The right panels are for $\text{Am}=10$ and the left panels are for $\text{Am}=100$. For each panel, we compare the fiducial box with $L_z=9 H_0$ (runs PL-ZNF-Am10 and PL-ZNF-Am100) with the taller box with $L_z=12 H_0$ (runs PL-ZNF-Am10-Lz12 and PL-ZNF-Am100-Lz12).}
    \label{fig_boxsize}
\end{figure*}

Another important test, especially for dynamo problems, is the dependence of the dynamo properties on the vertical box size. Our fiducial simulations are run with $L_z=9H_0$, and so we extend the box size to $L_z=12 H_0$, keeping the same resolution per $H_0$. Using {\tt PLUTO}, we first re-computed the hydrodynamical states with the extended box assuming $T^4$ cooling. We found very similar box-averaged quantities between $L_z=9H_0$ and $L_z=12H_0$ (Table \ref{table2}). We then performed two MHD runs with ambipolar diffusion, having $\text{Am}=10$ and $\text{Am}=100$ (using {\tt PLUTO}). To avoid the long kinematic phase, we introduced a strong magnetic perturbation at the beginning of the simulation of the form $B_y=0.1\exp{(-z^2/8H_0^2)}$. Figure \ref{fig_boxsize} (top panels) shows the vertical profiles of $\overline{B}_x$ and $\overline{B}_y$ for $L_z=9H_0$ and $L_z=12H_0$, averaged during the saturated state. Clearly, for both $\text{Am}=10$ and $\text{Am}=100$, the magnetic profiles are almost independent of vertical box size. The bottom panels of Fig.~\ref{fig_boxsize} show the ideal part of the Poynting flux as a function of $z$ for both $L_z=9H_0$ and $L_z=12H_0$. There is again little difference between the two, indicating that the amount of magnetic energy that leaves the disc surface is almost identical. We also checked that the EMF profiles are not affected by increasing $L_z$. All of these results suggest that the dynamo properties are independent of vertical box size.

\section{Discussion and Conclusions}

In conclusion, we conducted a series of 3D MHD shearing-box simulations of self-gravitating discs, stratified in the vertical direction and with zero net flux, to understand whether gravitational instability is able to generate and sustain a large-scale magnetic field. The physical and numerical setup is similar to that used in \citetalias{riols19b}. The main difference is that we examined the effect of ambipolar diffusion (rather than Ohmic dissipation) on the dynamo, a non-ideal effect that is more likely to impact the GI-unstable regions of poorly ionised protoplanetary discs. We examined the fundamental properties of the dynamo and its saturation for a wide range of ambipolar Elsasser numbers (Am). We also studied the dependence of the dynamo on the cooling law, resolution and box size. We obtained the following results: 
\begin{enumerate}
    \item The GI dynamo in the presence of ambipolar diffusion is optimal for ambipolar Elsasser numbers between $\text{Am}=30$ and $100$. In that regime, the magnetic field reaches thermal strength, with $\beta \sim 1$. The field is mainly large scale ($\sim$ box size in the horizontal direction) and dominated by the midplane toroidal component. At larger Am the magnetic energy weakens, which suggests that the dynamo is `slow'.
     These results are reminiscent of the case with Ohmic dissipation \citepalias{riols19b}. At small Am ($\text{Am}<20)$, the field strength decreases and the magnetic to thermal energy ratio scales as: 
     \begin{equation}
        E_m/E_T \sim 0.02\, \text{Am}^{1.15}
    \end{equation}
     Therefore the field strength is expected to vary according to the following power law: 
    \begin{equation}
       B \sim 0.15\, \text{Am}^{0.57} (\Sigma H)^{1/2} \Omega
    \end{equation}

    \item The mechanism of field generation is similar to that described in \citetalias{riols19b} and relies on the vertical circulations (poloidal `rolls') associated with spiral density waves. We used a new phase-folding procedure that made it especially clear to see these characteristic motions and the consequent field topology . 
    
    \item The saturation level of the dynamo is independent of the cooling law, except that we obtained more variability (in particular, thermal cycles) when a linear cooling law is employed. The origin of these cycles is still not well understood. 
    
    \item Finally, the dynamo is independent of the code used and robust to various numerical details (e.g. vertical box size, EMF reconstruction algorithm). We found that resolution does not affect the results provided that Am is sufficiently small ($\text{Am} \lesssim 100$). However, in the ideal-MHD limit (without explicit diffusion), the results are not converged with resolution, suggesting that diffusion plays a crucial role in the amplification of the field. The behaviour of the dynamo as \text{Am} tends to infinity remains difficult to predict.  
\end{enumerate}

The dynamo process that we study in this paper has potential direct implications for magnetic-field generation in the outer regions (typical beyond 20 AU) of young and massive protoplanetary discs. It is expected that in these regions the Class 0/I discs are subject to ambipolar diffusion and self-gravity, though the values of Am in the midplane of such discs are highly uncertain. Indeed, detailed ionisation maps of young Class 0/I disc have not been produced in the literature, on contrast to classical T Tauri discs. Nevertheless, the chemical models of \citet{masson16} suggest a range of Am between 0.02 and 20 in the outer regions ($R>50~{\rm au}$) of nascent protostellar discs during the protostellar core collapse (note that their definition of Am is different from ours and so we have converted their Am to the one we use in this paper). According to our Fig.~\ref{fig_saturation}, this lies in the regime of weak dynamo, but the range of expected values for the magnetic-field strength remains fairly broad. In the best case ($\text{Am}\sim 20$), the magnetic energy is roughly half of the thermal energy, which implies field strengths of a few tens of mG at $30~{\rm au}$ \citepalias[see discussion of ][for details of this estimation]{riols19b}. This is comparable with values of the field estimated from meteoritic samples, although these values are generally given at $R \lesssim 10~{\rm au}$ \citep{fu20}. Overall, given the uncertainty over Am it is currently a challenge to predict the intensity of the magnetic field generated by GI in Class 0/I discs. \\

Our work is also relevant to understanding the generation of magnetic fields in AGN discs. First, given their small thickness, AGN discs are susceptible to gravitational instability \citep{menou01,goodman03,lodato07} beyond ${\sim}0.01~{\rm pc}$, and therefore could be subject to the GI dynamo. For the relatively low densities in these regions, ambipolar diffusion can be more important than resistive diffusion, but for reasonable mass accretion rates models predict that $\text{Am} \gg 100$ \citep{menou01}. Our simulations cannot probe this regime but suggest that the dynamo would not be optimal. Such values of Am remain poorly constrained and depend on the efficiency of thermal ionisation and photo-ionisation. It can be also influenced by the geometry of the disc in the case it is warped. In any case, AGN (as well as protoplanetary discs) are believed to be threaded by a large-scale vertical magnetic field \citep{prior19}, which could significantly alter the dynamo properties. Future work is needed to understand the impact of this vertical field. This will be the focus of paper II. 

Finally it is worth mentioning that our work might be applied to galactic dynamos. Indeed radio observations revealed that some spiral galaxies  (M81,  M51 and  M33) are dominated by bisymmetric-spiral magnetic fields. These magnetic spiral patterns are strongly correlated with the large scale galactic spiral arms. Our result suggest that spiral density waves can directly produce magnetic fields in these objects as well as the observed alignment between magnetic and density spiral patterns (see Fig.~\ref{fig_Am10_xy}). This scenario could compete with to the standard mean field theory which is usually invoked to explain the dynamo in galactic discs \citep{chiba90,moss97,elstner00,rudiger04,gressel08}.

\label{sec_conclusions}

\section*{Acknowledgments}

AR and GL have received funding from the European Research Council (ERC) under the European Union’s Horizon 2020 research and innovation programme (Grant agreement No. 815559 (MHDiscs)). High-performance computing resources were provided by the GRICAD infrastructure (https://gricad.univ-grenoble-alpes.fr), which is supported by Grenoble research communities; and the PICSciE-OIT TIGRESS High Performance Computing Center and Visualization Laboratory at Princeton University.

\section*{Data availability}
The data underlying this article will be shared on reasonable request to the corresponding author.

\bibliographystyle{mnras}
\bibliography{refs} 

\appendix 

\section{Test of {\tt Athena++} implementation}
\label{appendixA}
\begin{figure}
    \centering
    \includegraphics[width=.5\textwidth]{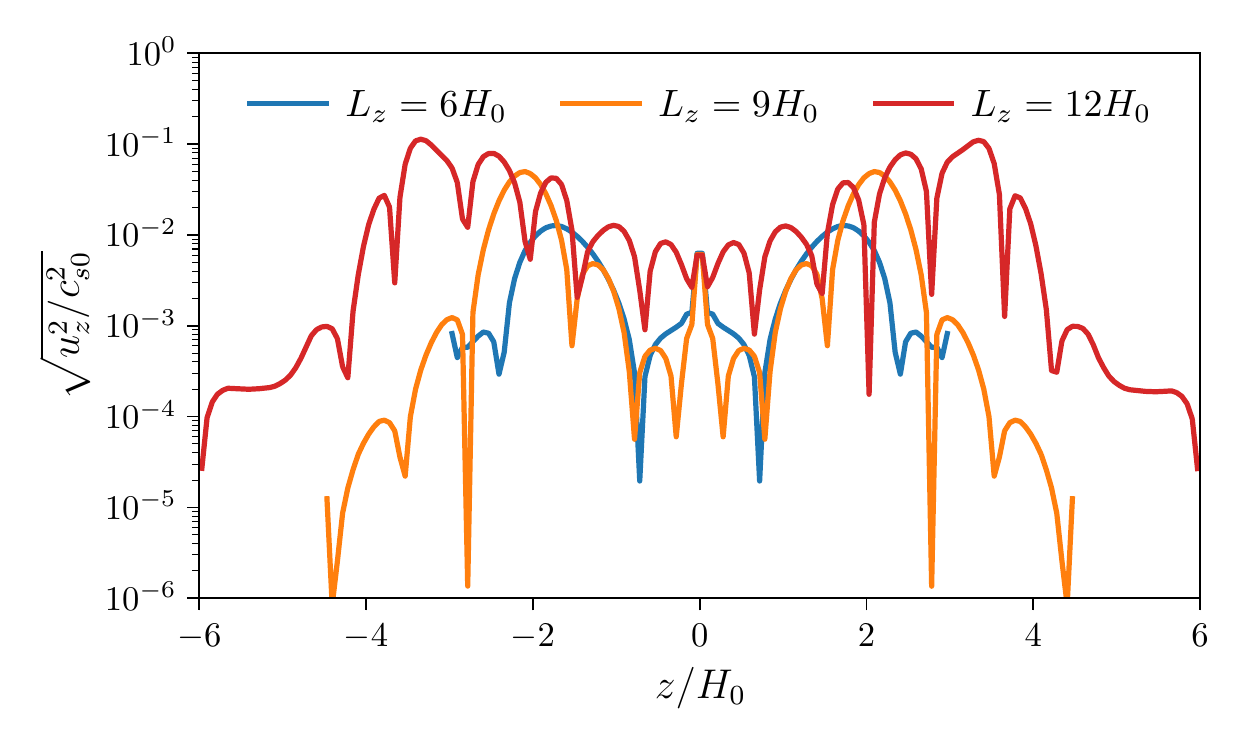}
    \caption{Vertical velocity perturbation for different vertical box sizes after evolving the background disk equilibrium for $t=100\Omega^{-1}$ using {\tt Athena++}.}
    \label{fig:Athena_test_1}
\end{figure}
\begin{figure}
    \centering
    \includegraphics[width=.4\textwidth]{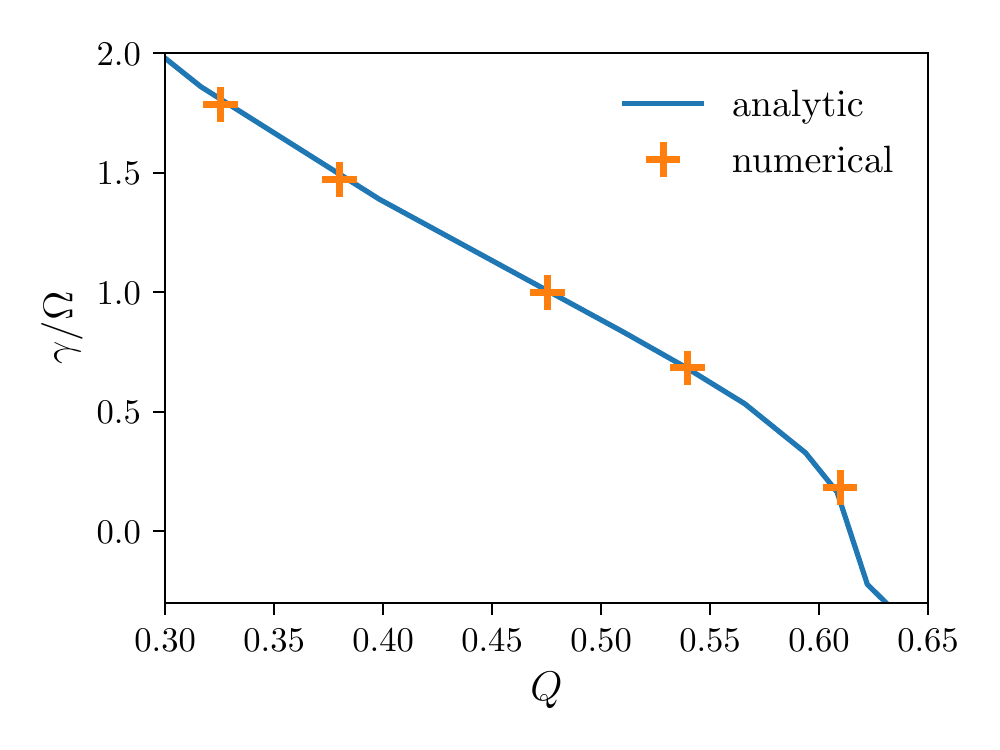}
    \caption{Growth rate of the most unstable axisymmetric mode with $k_x=2\pi/5H_0$ for an isothermal disk with different $Q$. The blue curve is the analytic result from \citet{riols17b} and orange crosses are measured from {\tt Athena++} simulations.}
    \label{fig:Athena_test_2}
\end{figure}

In this appendix we present two tests of our numerical setup in {\tt Athena++} to verify that self gravity is implemented correctly. These tests are similar to those in Appendix A and B of \citet{riols17b}, which were used to test self gravity in {\tt PLUTO}.

We first examine the stability of the background hydrostatic equilibrium. We initialize a un-magnetized disk in hydrostatic equilibrium with uniform temperature and $Q=1$. The setup (including vertical resolution) is identical to the initial condition of our hydrodynamic simulations, except that we do not introduce any initial perturbation and we use a much smaller domain in $x$ and $y$ so that the disc is stable against gravitational instability. We evolve this equilibrium state for 100$\Omega^{-1}$ for different box sizes $L_z$.
The amplitude of fluctuations in the velocity at 100$\Omega^{-1}$, shown in Fig.~\ref{fig:Athena_test_1}, increases slightly for a taller box, but remains well below the sound speed in all cases. The $L_z=6H_0$ result is similar to Figure A2 of \citet{riols17b}. Excluding the floor density from the gravitational force is necessary for a tall box, otherwise the velocity perturbation is much larger in the atmosphere.

Our second test measures the linear growth rate of the fastest growing axisymmetric mode and compares it with analytic result from \citet{riols17b}. For simplicity, we consider a disc with an isothermal equation of state. Figure \ref{fig:Athena_test_2} shows the growth rate of modes with $k_x=2\pi/5H_0$ for different $Q$; the growth rates measured from simulations agree very well with analytic predictions.

\section{Dependence on the EMF reconstruction scheme}

\begin{figure}
    \centering
    \includegraphics[width=\columnwidth]{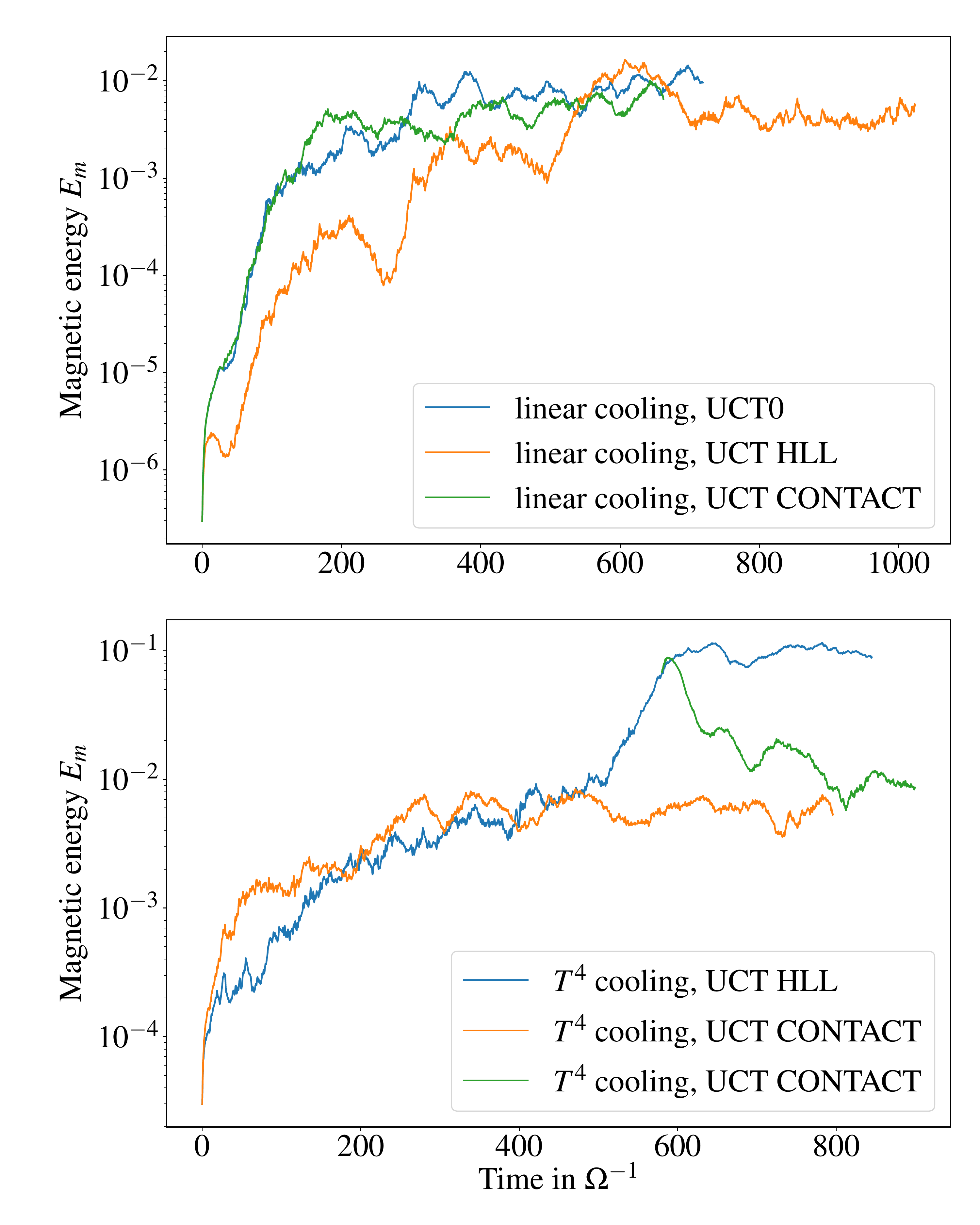}
    \caption{Time-evolution of magnetic energy for different EMF reconstruction schemes and cooling laws (top: linear cooling, bottom: $T^4$ cooling). In the bottom panel the orange line corresponds to a simulation with UCT CONTACT initialised from a seed field, while the green curve has the same reconstruction scheme but starts using a state from the UCT HLL run.}
    \label{fig_reconst_EMF}
\end{figure}

We study in this appendix the dependence of the dynamo behaviour on the reconstruction scheme of the EMF. In {\tt PLUTO} and {\tt Athena++}, staggered components of the magnetic field are evolved as surface-averaged quantities and are updated using Stokes' theorem. This requires constructing an averaged EMF along the face edges using a reconstruction or averaging technique from the face centres. Different techniques are employed, all of them having different levels of numerical diffusion. The UCT HLL method uses a two-dimensional Riemann solver based on a four-state HLL flux function \citep[see][]{zanna03}. It is generally the most diffusive technique. The UCT0 and UCT CONTACT employs the face-to-edge integration procedures proposed by \citet{gardiner05}, where electromotive force derivatives are averaged from neighbouring zones (UCT0) or selected according to the sign of the contact mode (UCT CONTACT). These methods are known to have reduced diffusion. Figure \ref{fig_reconst_EMF} shows the evolution of magnetic energy computed in {\tt PLUTO} without explicit diffusion using different EMF reconstruction schemes. In case of linear cooling, all methods seem to converge toward the same state with $E_m \lesssim 0.01$. However, in the case of $T^4$ cooling, we see an important gap between the UCT HLL (blue line) and UCT CONTACT (orange line). In particular, the UCT HLL scheme seems to produce a much stronger field, in agreement with the fact that the GI dynamo is boosted in the presence of a small amount of dissipation. 
{UCT0 is not part of this test because the method appears to be numerically unstable for this problem. This is somewhat expected, given that this reconstruction scheme is the one which is the least dissipative, and which is known to produce spurious oscillations even in the linear regime \citep{gardiner05}.}
The green line in Fig.~\ref{fig_reconst_EMF} corresponds to a simulation using the UCT CONTACT method and starting from the strong dynamo state obtained with UCT HLL. We see that the magnetic energy decays rapidly to the state corresponding to the orange curve. This rules out the existence of a bi-stable dynamo state and indicates that the non-linear saturation of the ideal simulation is dependent on the reconstruction scheme (and in particular sensitive to the level of  numerical diffusion induced by these scheme).

\begin{figure}
    \centering
    \includegraphics[width=\columnwidth]{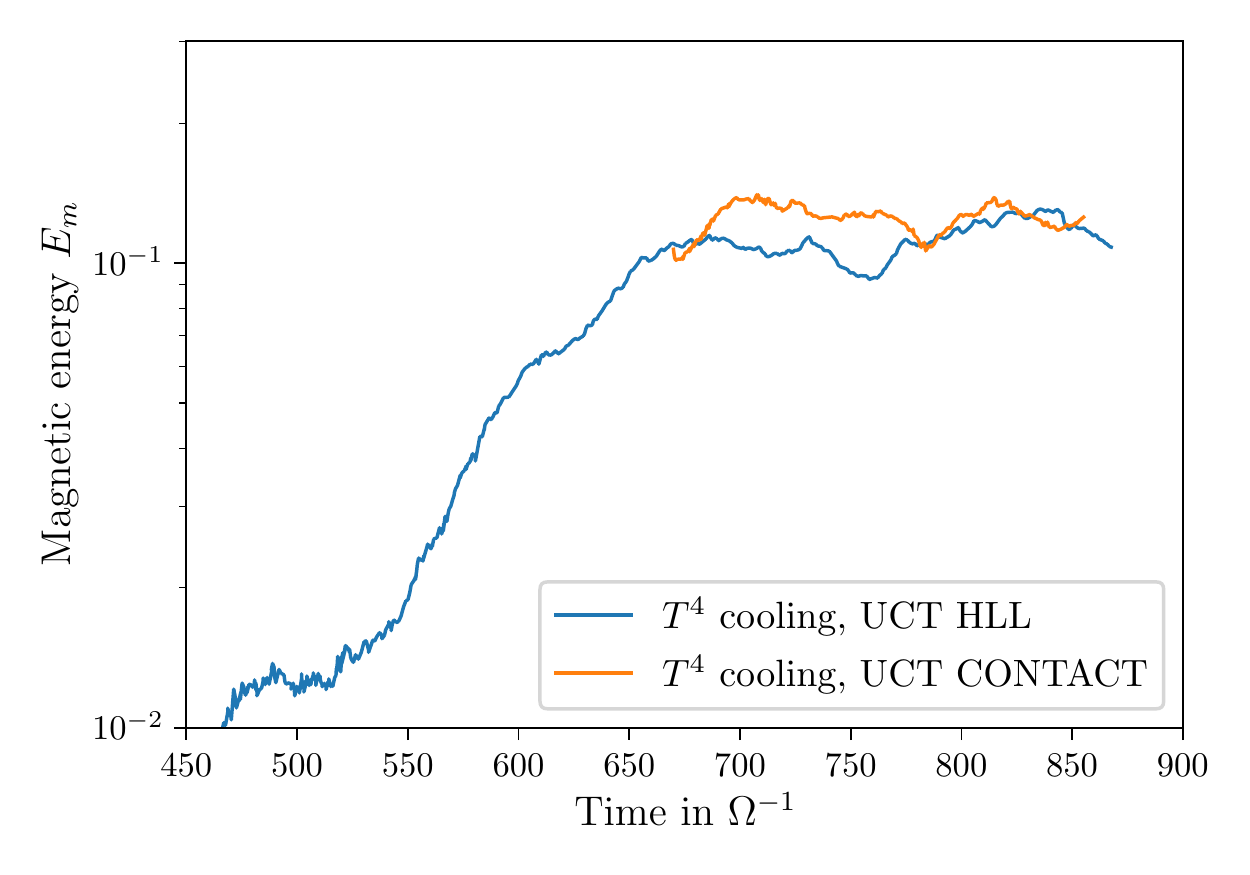}
    \caption{Same as the bottom panel of Figure \ref{fig_reconst_EMF}, but with ${\rm Am}=100$. There is no longer a significant difference between UCT HLL and UCT CONTACT.}
    \label{fig_reconst_EMF_Am100}
\end{figure}

{The result is expected to be much less sensitive to the reconstruction scheme when ambipolar diffusion is included, since in this case the dissipation of field at small scale is no longer dominated by numerical dissipation. Indeed, at ${\rm Am}=100$, the magnetic energy saturates to the same value for both UCT HLL and UCT CONTACT (Figure \ref{fig_reconst_EMF_Am100}).}


\label{appendixB}

\section{Dynamo with ionised layer}
\label{appendixC}
\begin{figure*}
    \centering
    \includegraphics[width=\textwidth]{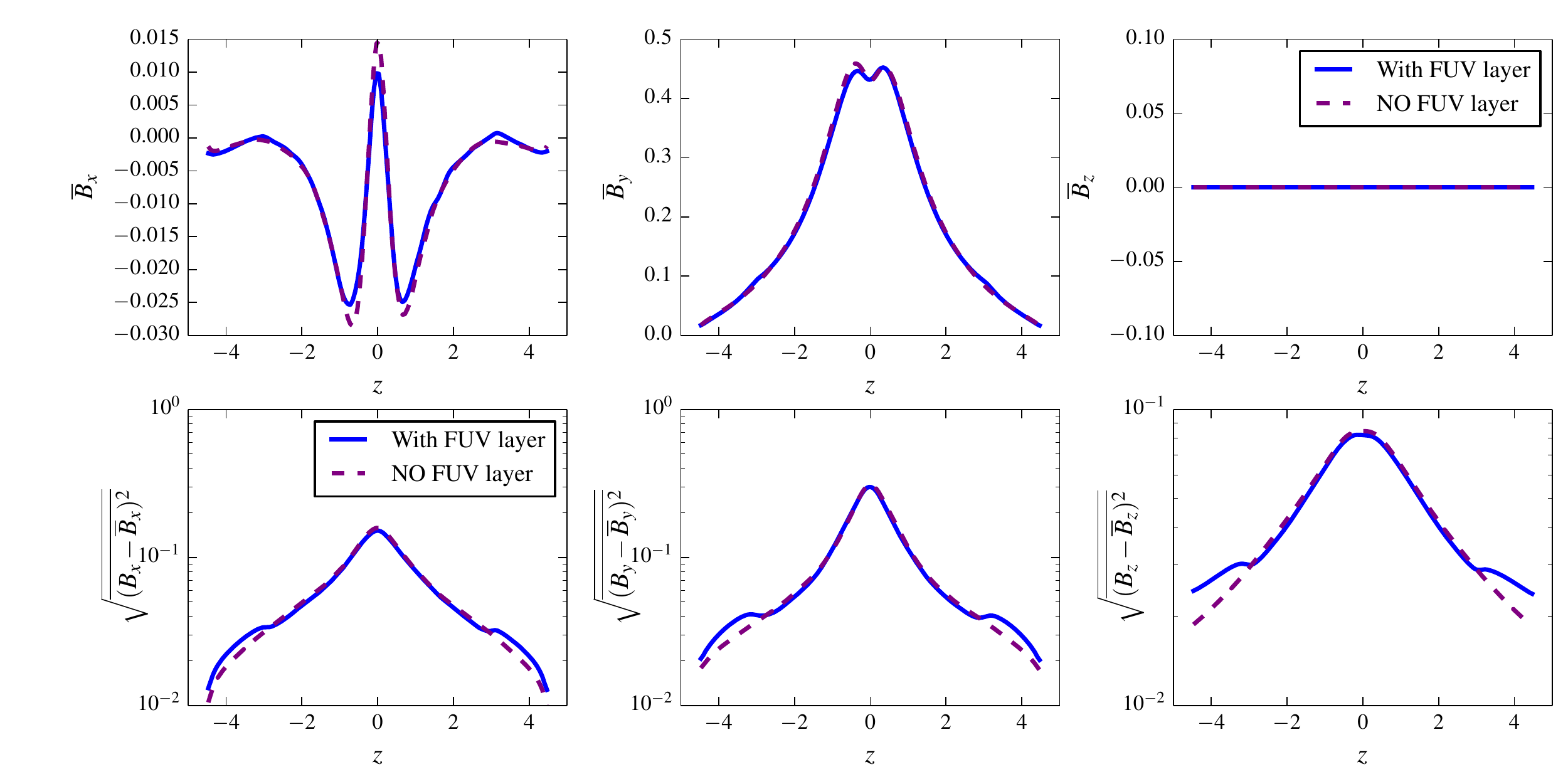}
    \caption{Top: vertical profile of mean magnetic field components in the radial, azimuthal and vertical directions. Bottom: root-mean-square magnetic fluctuations in each direction, where fluctuations are defined as $\mathbf{B}-\overline{\mathbf{B}}$. The blue/solid line is for the simulation with an ionised layer and $\text{Am}=10$ in the midplane. The purple/dashed line is for the simulation with uniform $\text{Am}=10$.}
\label{fig_FUV}
\end{figure*}

So far, all of our simulations have employed a simple prescription with uniform ion density and Elasasser number $\text{Am}$ in the box. However it is expected that, when the star is formed (Class I discs), it radiates FUV and potentially ionises the surface of the disc. We then test the dynamo behaviour when such an ionised layer is present. This allows us to check also the robustness of the dynamo regarding the physical environment present above the disc surface and near the box vertical boundary. To simplify the problem, we use the same model as in \citet{simon15}, which captures the essential physics of disc ionisation, namely, the effect of FUV in the disc corona. In this model,  Am is constant in the midplane (we chose $\text{Am}=10$) and increases abruptly up to a certain height $z_{io}$. This height corresponds to the altitude at which FUV can penetrate, i.e., when the horizontally-averaged density integrated from the vertical boundary, $\Sigma_i(z)$,  becomes larger than $\Sigma_{ic}=0.01~{\rm g}~{\rm cm}^{-2}$. The profile of Am is then given by:
\begin{equation}
\text{Am} = \text{Am}_{0} +3.3\times 10^{12}x_i(z)\,\dfrac{\rho}{\rho_0} \,\,(R_0/1 ~\text{au})^{-5/4}, 
\end{equation}
where $\text{Am}_{0}=10$ is the  constant midplane value, $R_0=50~{\rm au}$ is the radius of interest,  and 
\begin{equation}
x_i(z)=10^{-1}\exp\left[-\left(\dfrac{\Sigma_i(z)}{\Sigma_{ic}}\right)^4\right].
\end{equation}
is the ionisation fraction. Note that the ionisation fraction considered in the corona is much higher than what one might expect in a real protoplanetary disc (typically $x_i\sim 10^{-5}$), but this value ensures that we are almost in the ideal regime with $\text{Am} \ggg 10$. To convert numerical $\Sigma_i$  into ${\rm g}~{\rm cm}^{-2}$, we have used a typical radial profile of Class 0--I discs with $\Sigma = 1500 \, (R/1~{\rm au})^{-1}~{\rm g}~{\rm cm}^{2}$. Using such a prescription, the transition between the disc and the ionised layer takes place at $z=3H_0$. Above this height, Am reaches values ${\sim}10^5$ to ${\sim}10^6$. 

The simulation with an ionised layer is initialised using a state taken from run PL-ZNF-Am10 and is run for a time $200\Omega^{-1}$. Figure \ref{fig_FUV} (top) shows the vertical profiles of the mean magnetic field components averaged in time over the saturated state, for the case with an ionised layer and the case without (run PL-ZNF-Am10). We see that the profiles are almost identical in both cases, suggesting that the large-scale GI dynamo is not altered by the presence of an FUV layer. We also plot in the bottom panels the vertical profiles of the root-mean-square magnetic fluctuations, which provides an idea of the vertical distribution of the turbulent field. Again there is no major difference between the case with and without FUV layer, except that turbulent magnetic fluctuations are slightly stronger for $z>3H_0$ in the case with an FUV layer.

\label{lastpage}
\end{document}